%% file: main.tex
\documentclass[5p]{elsarticle} 

\usepackage{algpseudocode}
\usepackage{multirow}
\usepackage{footnote}
\usepackage{tabularx}
\usepackage{booktabs}
\usepackage{graphicx}
\usepackage{flushend}
\usepackage{amsmath}
\usepackage{tablefootnote}
\usepackage{hyperref}
\usepackage{enumitem}
\usepackage[toc,page]{appendix}
\usepackage[dvipsnames]{xcolor}
\usepackage{comment}
\usepackage[utf8]{inputenc}
\usepackage[linesnumbered,ruled]{algorithm2e}

\newcommand{\term}[1]{\begin{small}\texttt{#1}\end{small}}
\newtheorem{defn}{Definition}

\newcounter{example}
\newenvironment{example}[1][]{\refstepcounter{example}\par\medskip
   \noindent \textbf{Example~\theexample. #1} \rmfamily}{\medskip}
\usepackage{textcomp}
\newcommand{\textapprox}{\raisebox{0.5ex}{\texttildelow}}

\begin{document}

\newcommand{\sr}[1]{{\textcolor{red}{SR: #1}}}
\newcommand{\GW}[1]{{\color{purple}{GW: #1}}}

\title{Negative Statements Considered Useful}

\author[add1]{Hiba Arnaout\corref{cor1}}
\ead{harnaout@mpi-inf.mpg.de}
\author[add1]{Simon Razniewski}
\ead{srazniew@mpi-inf.mpg.de}
\author[add1]{Gerhard Weikum}
\ead{weikum@mpi-inf.mpg.de}
 \author[add2]{Jeff Z. Pan}
\ead{http://knowledge-representation.org/j.z.pan/}

\cortext[cor1]{Corresponding author}
\address[add1]{Max Planck Institute for Informatics, Saarland Informatics Campus, Saarbr{\"u}cken 66123, Germany}
\address[add2]{School of Informatics, The University of Edinburgh, Informatics Forum, Edinburgh EH8 9AB, Scotland}

\begin{abstract}

Knowledge bases (KBs)
about notable entities and their properties
are an important asset in applications such as search, question answering and dialogue. 
All popular KBs capture virtually only positive statements, 
and abstain from taking any stance on statements not 
stored in the KB.
This paper makes the case for explicitly stating salient statements 
that
do \textit{not} hold.
Negative statements are useful to overcome
 limitations of question answering systems that are mainly geared for positive questions;
 they  can also contribute to informative summaries of entities.
 Due to the abundance of such invalid statements, any effort to compile them needs to address ranking by saliency. We present a statistical inference method for 
 compiling and ranking negative statements, based on expectations from positive statements of related  entities
 in peer groups. 
 Experimental results, with a variety of datasets,
 show that 
 the method can effectively
 discover notable negative statements,
 and extrinsic studies underline their usefulness for
 entity summarization.
 Datasets and code 
 are released as resources for further research.

\end{abstract}

\begin{keyword}
knowledge bases, negative knowledge, 
information extraction, 
statistical inference,
ranking
\end{keyword}

\maketitle    

\input{introduction}
\input{related}

\input{model}
\input{peer}
\input{orderedpeer}
\input{conditional}
\input{experiments}
\input{extrinsic}
\input{resources}

\input{discussion}

\input{conclusion}

\bibliographystyle{splncs04}
\bibliography{refs}

\end{document}

%% file: introduction.tex
\section{Introduction}
\label{sec:intro}
\noindent
\textbf{Motivation and Problem.\ }
Structured knowledge is crucial in a range of applications like question answering, dialogue agents, and recommendation systems. The required knowledge is usually stored in KBs, and recent years have seen a rise of interest in KB construction, querying and maintenance, with notable projects being Wikidata~\cite{WD}, DBpedia~\cite{DBPEDIA}, Yago~\cite{YAGO}, or the Google Knowledge Graph~\cite{GKG}. These KBs store positive statements such as \textit{``Ren\'{e}e Zellweger won the 2020 Oscar for the best actress''}, and are a key asset for many knowledge-intensive AI applications.

A major limitation of all these KBs is their inability to deal with negative information~\cite{FHPPW2006}. At present, most major KBs only contain positive statements, whereas statements such as that \textit{``Tom Cruise did not win an Oscar''} could only be inferred with the major assumption that the KB is complete - the so-called \textit{closed-world assumption} (CWA). Yet as KBs are only pragmatic collections of positive statements, the CWA is not realistic to assume, and there remains uncertainty whether statements not contained in a KBs are false, or truth is merely unknown to the KB. 

Not being able to formally distinguish whether a statement is false or unknown poses challenges in a variety of applications. In medicine, for instance, it is important to distinguish between knowing about the absence of a biochemical reaction between substances, and not knowing about its existence at all. In corporate integrity, it is important to know whether a person was never employed by a certain competitor, while in anti-corruption investigations, absence of family relations needs to be ascertained. In data science and machine learning, on-the-spot counterexamples are important to ensure the correctness of learned extraction patterns and associations.

\noindent
\textbf{State of the Art and its Limitations.\ }
Absence of explicit negative knowledge has consequences for usage of KBs: for instance, today's \textit{question answering} (QA) systems are well
geared for positive questions, and questions where exactly one answer should be returned (e.g., quiz questions or reading comprehension tasks) \cite{Fader2014,WIKIQA}. In contrast, for answering negative questions like \emph{``Actors without Oscars''}, QA systems lack a data basis. Similarly,
they struggle with positive questions that have no answer, like \emph{``Children of Emmanuel Macron''},
too often still returning a best-effort answer even if
it is incorrect. Materialized negative information would allow a better treatment of both cases.

\noindent
\textbf{Approach and Contribution.\ }
In this paper, we make the case that important negative knowledge should be explicitly materialized. We motivate this selective materialization with the challenge of overseeing a near-infinite space of false statements,
 and with the importance of explicit negation in search and question answering. 

We consider three classes of negative statements: (i) groun\-ded negative statements \textit{``Tom Cruise is not a British citizen''}, (ii) conditional negative statements \textit{``Tom Cruise has not won an award from the Oscar categories''} and (iii) universal negative statements \textit{``Tom Cruise is not member of any political party''}. In a nutshell, given a KB and an entity e, we select highly related entities to e (we call them \textit{peers}). We then use these peers to derive positive expectations about e, where the absence of these expectations might be interesting for e. In this approach, we are assuming completeness within a group of peers. More precisely, if the KB does not mention the \textit{Nobel Prize in Physics} as an award won by \textit{Stephen Hawking}, but does mention it for at least one of his peers, it is assumed to be false for \textit{Hawking}, and not a missing statement. This is followed by a ranking step where we use predicate and object prominence, frequency, and textual context in a  learning-to-rank model.

The salient contributions of this paper are:

\begin{enumerate}
    \item We make the first comprehensive case for materializing \textit{useful} negative statements, and formalize important classes of such statements.
    \item We present a judiciously designed method for collecting and ranking negative statements based on knowledge about related entities.
    \item We show the usefulness of our models in use cases like entity summarization, decision support, and question answering.\\
Experimental datasets and code are released as resources for further research\footnote{\url{https://www.mpi-inf.mpg.de/departments/databases-and-information-systems/research/knowledge-base-recall/interesting-negations-in-kbs/}}.
\end{enumerate}

The present article extends the earlier conference publication~\cite{negationakbc} in several directions:

\begin{enumerate}
    \item We extend the statistical inference to ordered sets of related entities, thereby removing the need to select a single peer set, and obtaining finer-grained contextualizations of negative statements (Section~\ref{sec:temporal}); 
    \item To bridge the gap between overly fine-grained groun\-ded negative statements and coarse universal negative statements, 
    we introduce a third notion of negative statement, \emph{conditional negative statements}, and show how to compute them post-hoc (Section~\ref{sec:restricted});
    \item We evaluate the value of negative statements in an additional use case, with hotels from Booking.com (Section~\ref{sec:extrinsicevaluation}).
\end{enumerate}

%% file: related.tex
\section{State of the Art}
\label{sec:related}


\subsection{Negation in Existing Knowledge Bases}
\label{sec:existing}




\noindent
\textbf{Deleted Statements.\ }
Statements that were once part of a KB but got subsequently deleted are promising candidates for negative information~\cite{edithistory2019}. As an example, we studied deleted statements between two Wikidata versions from 1/2017 and 1/2018, focusing in particular on statements for people (close to 0.5m deleted statements). On a random sample of 1k deleted statements, we found that over 82\% were just caused by ontology modifications, granularity changes, rewordings, or prefix modifications. 
Another 15\% were statements that were actually restored a year later, so presumably reflected erroneous deletions. The remaining 3\% represented actual negation, yet we found them to be rarely noteworthy, i.e., presenting mostly things like corrections of birth dates or location updates reflecting geopolitical changes.

In Wikidata, erroneous changes can also be directly recorded via the deprecated rank feature~\cite{MKGGB2018}. Yet again we found that this mostly relates to errors coming from various import sources, and did not concern the active collection of interesting negations, as advocated in this article.

\noindent
\textbf{Count and Negated Predicates.\ }
Another way of expressing negation is via counts matching with instances, for instance, storing 5 children statements for \textit{Trump} and numerical statement \term{(number of children; 5)} allow to infer that anyone else is not a child of \textit{Trump}. Yet while such count predicates exist in popular KBs, none of them has a formal way of dealing with these, especially concerning linking them to instance-based predicates~\cite{ghoshSWJ}.

Moreover, some KBs contain relations that carry a negative meaning. For example, DBpedia has predicates like \emph{carrier never available} (for phones), 
or \emph{never exceed alt} (for airplanes), 
Knowlife~\cite{ernst2015knowlife} contains medical predicates like \emph{is not caused by} 
and \emph{is not healed by}, 
and Wikidata contains \emph{does not have part} and 
\emph{different from}. 
Yet these present very specific pieces of knowledge, and do not generalize. 
Although there have been discussions to extend the Wikidata data model to allow generic opposites\footnote{\url{https://www.wikidata.org/wiki/Wikidata:Property_proposal/fails_compliance_with}}, these have not been worked out so far.

\noindent
\textbf{Wikidata No-Values.\ }
Wikidata can capture statements about \textit{universal absence} via the ``no-value'' symbol~\cite{erxleben2014introducing}. This allows KB editors to add a statement where the object is empty. For example, what we express as \term{$\neg \exists x$(Angela Merkel; child; x)}, the current version of Wikidata allows to be expressed as \term{(Angela Merkel; child; no-value)}\footnote{\url{https://www.wikidata.org/wiki/Q567}}. As of 8/2021, there exist 135k of such ``no-value'' statements, yet only used in narrow domains. For instance, 53\% of these statements come for just two properties \textit{country} (used almost exclusively for geographic features in Antarctica), and \textit{follows} (indicating that an artwork is not a sequel).


\subsection{Negation in Logics and Data Management}


Negation has a long history in logics and data management. Early database paradigms usually employed the closed-world assumption (CWA), i.e., assumed that all statements not stated to be true were false \cite{Reiter78}, \cite{ICWA}. On the Semantic Web and for KBs, in contrast, the open-world assumption (OWA) has become the standard. The OWA asserts that the truth of statements not stated explicitly is unknown. Both semantics represent somewhat extreme positions, as in practice it is neither conceivable that all statements not contained in a KB are false, nor is it useful to consider the truth of all of them as unknown, since in many cases statements not contained in KBs are indeed not there because they are known to be false~\cite{razniewskilimits}. Between these two assumptions, there is also the so-called local (partial) closed-world assumption~\cite{RPZ2010c}, where the open-world assumption is used in general, while the  closed-world assumption can be applied to some predicates (classes or properties).

In limited domains, logical rules and constraints, such as Description Logics \cite{logichandbook}, \cite{Calvanese2007} or OWL, can be used to derive negative statements. An example is the statement that every person has only one birth place, which allows to deduce with certainty that a given person who was born in \textit{France} was not born in \textit{Italy}. OWL also allows to explicitly assert negative statements \cite{mcguinness2004owl}, yet so far is predominantly used as ontology description language and for inferring intensional knowledge, not for extensional information (i.e., instances of classes and relations), with a few exceptions, like the rewriting based approach to instance retrieval for negated concepts, based on the notion of inconsistency-based first-order-rewritability~\cite{DuPa2015}. Different levels of negations and inconsistencies in Description Logic-based ontologies are proposed in a general framework~\cite{FHPPW2006}.

In~\cite{DBLP:journals/amai/AnalytiADP13,Analyti04negationand}, a thorough study on negative information in the Resource Description Framework (RDF) argues in favor of explicit negation. In particular, it makes the point that any knowledge representation formalism must be able to deal with \textit{informative} negative information, on top of informative positive information. The authors then propose ERDF (extended RDF), where an ERDF triple can be either positive or negative. The framework also distinguishes between two kinds of negation: weak (``she doesn't like snow'') and strong (``she dislikes snow''). The former is denoted using the \textapprox \   symbol, and the latter using the $\neg$ symbol.

The notion of \texttt{noValue} in RDF was introduced in~\cite{DBLP:books/aw/AbiteboulHV95}. It has been recently adapted in~\cite{DBLP:conf/semweb/DarariPN15} for representing no-value information in RDF and incorporating such information into query answering. The intuition behind it is to distinguish whether a result set of a SPARQL query is empty due to lack of information or actual negation.

The AMIE framework \cite{galarraga2017predicting} employed rule mining to predict the completeness of properties for given entities. This corresponds to learning  whether the CWA holds in a local part of the KB, inferring that all absent values for a subject-predicate pair are false. For our task, this could be a building block, but it does not address the inference of {\em useful} negative statements.

RuDiK~\cite{ortona2018rudik} is a rule mining system that can learn rules with negative atoms in rule heads (e.g., people born in \textit{Germany} cannot be
\textit{U.S.} president). This could be utilized towards predicting negative statements. 
Unfortunately, such rules predict way too many -- correct, but uninformative -- negative statements, essentially enumerating a huge set of people who are not \textit{U.S.} presidents. 
The same work also proposed a precision-oriented variant of the CWA that assumes negation only if subject and object are connected by at least one other relation. Unfortunately, this condition is rarely met in interesting cases. For instance, most of the negative statements in Table~\ref{tbl:may:einstein} have alternative connections between subject and object in Wikidata.

\subsection{Related Areas}

\noindent
\textbf{\small Linguistics and Textual Information Extraction (IE).\ } Negation is an important feature of human language \cite{Morante2012}. While there exists a variety of ways to express negation, state-of-the-art methods are able to detect quite reliably whether a segment of text is negated or not \cite{extendingnegex}, \cite{wu2014}.  There is also work on using knowledge graphs to help detect false statements in texts, such as news~\cite{PPLL+2018}. 

A body of work targets negation in medical data and health records. In \cite{cruzdiaz}, a supervised system for detecting negation, speculation and their scope in biomedical data is developed, based on the annotated BioScope corpus \cite{bioscope}.
In \cite{Goldin03learningto}, the focus is on negations via the keyword ``not''. The challenge here is the right scoping, e.g., ``Examination could not be performed due to the Aphasia'' does not negate the medical observation that the patient has Aphasia.
In \cite{KEhealth}, a rule-based approach based on NegEx~\cite{CHAPMAN}, and a vocabulary-based approach for prefix detection were introduced.
PreNex \cite{PRENEX} also deals with negation prefixes. The authors propose to break terms into prefixes and root words to identify this kind of negation. They rely on a pattern matching approach over medical documents.

In \cite{AKB}, an anti-knowledge base containing negations is mined from Wikipedia change logs, with the focus however being again on factual mistakes, and precision, not interestingness, is employed as main evaluation metric. In~\cite{CSKB}, the focus is to obtain meaningful negative samples for augmenting commonsense KBs.
We explore text extraction in more details in the proposed \emph{pattern-based query log extraction} method in our earlier conference publication~\cite{negationakbc}.

\noindent
\textbf{Statistical Inference and KB Completion.\ } As text extraction often has limitations, data mining and machine learning are frequently used on top of extracted or user-built KBs, in order to detect interesting patterns in existing data, or in order to predict statements not yet contained in a KB. There exist at least three popular approaches, rule mining, tensor factorization, and vector space embeddings \cite{KGembsurvey}. Rule mining is an established, interpretable technique for pattern discovery in structured data, and has been successfully applied to KBs for instance by the AMIE system \cite{AMIE3}. Tensor factorization and vector space embeddings are latent models, i.e., they discover hidden commonalities by learning low-dimensional feature vectors \cite{global2014}. To date, all these approaches only discover positive statements. On the other hand, if one considers logical entailments as a means to enhance such rule mining and latent model based approaches, such as in an iterative manner~\cite{WPKD2020}, negative statements in theory can be discovered with the help of disjoint axioms; however, the quality of knowledge graph completion methods still have room for improvement. Recently, an inference model has been proposed to build a knowledge graph with commonsense contradictions~\cite{ANION}, like ``Wearing a mask is seen as responsible'' is the contradiction of ``Not wearing a mask is seen as carefree''.

\noindent
\textbf{Ranking KB Statements.\ } In applications such as entity summarization over web-scale KBs, returned result sets are often very large. Ranking statements is a core task in managing access to KBs, with techniques often combining generative language-models for queries on weighted and labeled graphs~\cite{NAGA,Yahya2016,arnaoutjws2018}. In \cite{Bast}, the authors propose a variety of functions to rank values of type-like predicates. These algorithms include retrieving entity-related texts, binary classifiers with textual features, and counting word occurrences.  In \cite{huang2019contextual}, the focus is on identifying the informativeness of statements within the context of the query, by exploiting deep learning techniques. In this work, applications such as entity summarization returns a set of \textit{negative} statements. To assign each statement a relevance score, we use a mixture of the metrics that are usually used for ranking positive statements (e.g., frequency of property), and metrics that are specific for negative statements (e.g., unexpectedness).


%% file: model.tex
\section{Model}
\label{sec:formalization}


For the remainder we assume that a KB is a set of statements, each being a triple $(s;p;o)$ of subject $s$, property $p$ and object $o$.


Let $K^i$ be an (imaginary) ideal KB that perfectly represents reality, i.e., contains exactly those statements that hold in reality. Under the OWA, (practically) available KBs, $K^a$ contains correct statements, but may be incomplete, so the condition $K^a \subseteq K^i$ holds, but not the converse \cite{razniewski2011completeness}.
We distinguish two forms of negative statements.

\begin{defn}[Negative Statements] \mbox{ }
\begin{enumerate}[noitemsep,topsep=0pt,parsep=0pt,partopsep=0pt]
\item A grounded negative statement $\neg (s, p, o)$ is satisfied if $(s, p, o) \notin K^i$.
\item A universally  negative statement $\neg\exists o: (s, p, o)$ is satisfied if there exists no $o$ such that $(s; p; o) \in K^i$.
\end{enumerate} 
\end{defn}




An example of a grounded negative statement is that \textit{``Bruce Willis was not born in the U.S.''}, and is expressed as \term{$\neg$(Bruce Willis; born in; U.S.)}. An example of a universally negative statement is that \textit{``Leonardo DiCaprio has never been married''}, expressed as \term{$\neg\exists o:$(Leonardo DiC\-aprio; spouse; o)}. Both types of negative statements represent standard logical constructs, and could also be expressed in the OWL ontology language. Grounded negative statements could be expressed via negative property statements (e.g., {\small{\texttt{NegativeObjectPropertyAssertion }}{\small\texttt{(:born In :Bruce Willis :U.S.)}}}), while universally negative statements could be expressed via \texttt{ObjectAllValuesFrom} or \texttt{owl:complementOf} \cite{erxleben2014introducing} (e.g., {\small{\texttt{ ClassAssertion (ObjectAl\-lValuesFrom (:spouse owl:Nothing) :Leonardo Dicaprio)}}}). Without further constraints, for these classes of negative statements, checking that there is no conflict with a positive statement is trivial. In the presence of further constraints or entailment regimes, one could resort to (in)cons\-istency checking services \cite{logichandbook,Pan2017,gadwww}.


Yet compiling negative statements faces two other challenges.
First, being not in conflict with positive statements is a necessary but not a sufficient condition for correctness of negation, due to the OWA. In particular, $K^i$ is only a virtual construct, so methods to derive correct negative statements have to rely on the limited positive information contained in $K^a$, or utilize external evidence, e.g., from text. Second, the set of correct negative statements is near-infinite, especially for grounded negative statements. Thus, unlike for positive statements, negative statement construction/extraction needs a tight coupling with ranking methods.

\noindent
\textbf{Research Problem 1.\ }
Given an entity $e$, compile a ranked list of useful grounded negative and universally negative statements.

%% file: peer.tex
\section{Peer-based Statistical Inference}
\label{sec:inference}
We next present a method to derive useful negative statements by combining information from similar entities (``peers'') with supervised calibration of ranking heuristics. The idea is that peers that are similar to a given entity can give expectations on relevant statements that \textit{should} hold for the entity. For instance, several entities similar to the physicist \emph{Stephen Hawking} have won the \textit{Nobel in Physics}. We may thus conclude that him not winning this prize could be an especially useful statement. Yet related entities also share other traits, e.g., many famous physicists are \textit{U.S.} citizens, while \textit{Hawking} is \textit{British}. We thus need to devise ranking methods that take into account various clues such as frequency, importance, unexpectedness, etc.\\

\noindent
\textbf{Peer-based Candidate Retrieval.\ }
To scale the method to web-scale KBs, in the first stage, we compute a candidate set of negative statements using the CWA on certain parts of the KB, to be ranked in the second stage. Given a subject $e$, we proceed in three steps:
\begin{enumerate}[noitemsep,topsep=0pt,parsep=0pt,partopsep=0pt]
    \item \textit{Obtain peers:} We collect entities that set expectations for statements that $e$ could have, the so-called \textit{peer groups} of $e$. These groups can be based on (i) structured facets of the subject~\cite{RECOIN}, such as \textit{occupation, nationality}, or \emph{field of work} for people, or classes/types for other entities, (ii) graph-based measures such as distance or connectivity~\cite{ponza}, or (iii) entity embeddings such as TransE~\cite{transE}, possibly in combination with clustering, thus reflecting latent similarity. 
    \item \textit{Count statements:} We count the relative frequency of all predicate-object pairs (i.e., \term{(\_,p,o)}) and predicates (i.e., \term{(\_,p,\_)}) within the peer groups, and retain the maxima, if candidates occur in several groups. This way, statements are retained if they occur frequently in at least one of the possibly orthogonal peer groups.
    \item \textit{Subtract positives:} We remove those predicate-object pairs and predicates that exist for $e$.
\end{enumerate}
\  \\
Algorithm~\ref{alg:peer} shows the full procedure of the peer-based inference method. 
In line 2, groups of peers $P[]$ are selected based on some blackbox function \textit{peer\_groups}.
\begin{equation*}
P = [P_1, ... P_n] \text{, with } n>=1.
\end{equation*}
\noindent
Every group $P_i$ is a set of peers, defined as follows.
\begin{equation*}
P_i = \{pe_1, ..., pe_m\} \text{, with } m<=s.
\end{equation*}
Subsequently, for each peer group, it collects all the positive information that these peers have (line 6 and 7), and stores them as a list of candidate statements.
\begin{equation*}
candidates = \{st_1, ..., st_w\}.
\end{equation*}
A statement $st_j$ in $candidates$ is either a predicate P or a predicate-object pair PO.
After collecting information about the peers, the loop at line 10 iterates over the list of unique statements $ucandidates$, computes their relative frequency, and stores them in the final list of negations $N$. $N$ is a list of negation objects~\footnote{Here, object is meant as a data type and not a KB-triple object.}, where every object consists of a negation statement and its score.
\begin{equation*}
N = [(\neg st_1, sc_1), ..., (\neg st_r, sc_r)].
\end{equation*}
Across peer groups, it retains the maximum relative frequencies (hence, line 13), if a property or statement occurs across several. Before returning the top $k$ results as output (line 18), it subtracts those already possessed by entity $e$ (line 17).

\begin{algorithm*}[t!]
    \SetKwInOut{Input}{Input}
    \SetKwInOut{Output}{Output}
    \Input{\small knowledge base $\mathit{KB}$, entity $e$, peer collection function \textit{peer\_groups}, \small max. size of a peer group $s$, \small number of results $k$}
    \Output{ \small $k$-most frequent negative statement candidates for $e$}
        \textit{\textbf{P}}$[]$= \textit{peer\_groups}$(e, s)$ \Comment{List of peer group(s); Group $P_i$ at position $i$ is one group (set) with at most $s$ peers.}\\
         \textit{\textbf{N}}$[]$= $\emptyset$ \Comment{\small Ranked list of negative statements about $e$.}\\
    \For{$P_i$ $\in$ \textbf{P}}{ 
        $candidates$ = [] \Comment{\small Positive statements (i.e., predicate and predicate-object pairs) of $P_i$ members.}\\
        \For{$pe$ $\in$ $P_i$}{ 
            $candidates$+=$collectP(pe)$ \Comment{\small Collecting predicates that hold for one peer (pe).}\\
            $candidates$+=$collectPO(pe)$ \Comment{\small Collecting predicate-object pairs that hold for pe.}\\
        }
        $ucandidates$ = $unique(candidates)$\Comment{\small List of unique statements in $candidates$.}\\
        \For{$st$ $\in$ $ucandidates$}{
            $sc$ = $\frac{count(st, candidates)}{s}$ \Comment{\small sc computes how many peers share the statement st, normalized by $s$.}\\
            \If{$getnegation(N, st).score$ $<$ $sc$}{ $setscore(N, st, sc)$}  
        }
     }
        $N$-=$\mathit{inKB}(e,N)$ \Comment{\small Remove statements $e$ already has.}\\
        return $max(N,k)$
\caption{Peer-based candidate retrieval algorithm.}
\label{alg:peer}
\end{algorithm*}

\begin{example}
Consider the entity $e$=\textit{Brad Pitt}. Table \ref{tab:brad} shows a few examples of his peers and candidate negative statements.
We instantiate the peer group choice to be based on structured information, in particular, shared occupations with the subject, as in Recoin~\cite{RECOIN}. In Wikidata, \textit{Pitt} has 9 occupations, thus we would obtain 9 peer groups of entities sharing one of these with \textit{Pitt}.
\begin{equation*}
P = [\text{actors, film directors, ..., models}] \text{, with } n=9.
\end{equation*}
For readability, let us consider statements derived from only one of these peer groups, \emph{actor}. Let us assume 3 entities in that peer group.
\begin{equation*}
P_\text{actor} = \{\text{Russel Crowe, Tom Hanks,  Denzel Washington}\}
\end{equation*}
The list of negative candidates, $candidates$, are all the predicate and predicate-object pairs shown in the columns of the 3 actors. And in this particular example, $N$ is just $ucandidates$ with scores for only the \textit{actor} group.
\begin{equation*}
\begin{split}
N = [(\neg (\text{award; Oscar for Best Actor}), 1.0),\\  (\neg \exists x(\text{instagram; }x\text), 0.67),\\ 
(\neg (\text{citizen; New Zealand}), 0.33),\\ 
(\neg \exists x(\text{convicted; }x), 0.33),\\ 
(\neg \exists x(\text{child; }x), 1.0),\\  (\neg(\text{occupation; screenwriter}), 1.0),\\  (\neg(\text{citizen; U.S.}), 0.67)].
\end{split}
\end{equation*}
Candidates that \textit{hold} for \textit{Pitt} are then dropped.
\begin{equation*}
\begin{split}
N = [(\neg (\text{award; Oscar for Best Actor}), 1.0),\\  (\neg \exists x(\text{instagram; }x), 0.67),\\ 
(\neg (\text{citizen; New Zealand}), 0.33),\\ 
(\neg \exists x(\text{convicted;} x), 0.33),\\  (\neg(\text{occupation; screenwriter}), 1.0)].
\end{split}
\end{equation*}
 The top-k of the rest of candidates in $N$ are finally returned. The top-3 negative statements, for this example, are \term{$\neg$(award; Oscar for Best Actor)}, \term{$\neg$(occupation; screenwriter)}, and \term{$\neg \exists x$(instagram; x)}. 

The ``if'' statement at line 12 is only needed when multiple peer groups are considered for an entity. In the case where a negative statement is inferred from more than 1 group, only the version with the highest score is added to the final set. In the original (\textit{full}) example, \textit{Pitt} belongs to the group \textit{actor} and the group \textit{model}. The negation \term{$\neg$(occupation; screenwriter)} was inferred twice, once from each group, with a relative frequency of 0.9 from the \textit{actor} group and 0.2 from the \textit{model} group. We add the one with the higher score to the final set and disregard the other one. An alternative is to combine or compute the average of the scores across groups.

Note that without proper thresholding, the candidate set grows very quickly, for instance, if using only 30 peers, the candidate set for \textit{Pitt} on Wikidata is already about 1500 statements.
\end{example}\\

\begin{table*}
  \caption{Discovering candidate statements for \textit{Brad Pitt} from one peer group with 3 peers.}
  \label{tab:brad}
   \resizebox{\textwidth}{!}{\begin{tabular}{lll|l|l}
    \toprule
 \multicolumn{1}{c}{\bf{Russel Crowe}} &  \multicolumn{1}{c}{\bf{Tom Hanks}} &  \multicolumn{1}{c}{\bf{Denzel Washington}} &  \multicolumn{1}{|c}{\bf{Brad Pitt}} & \multicolumn{1}{|c}{\bf{Candidate statements}}\\
    \midrule
(award; Oscar for Best Actor) & (award; Oscar for Best Actor) & (award; Oscar for Best Actor) & (citizen; U.S.) & $\neg$(award; Oscar for Best Actor), 1.0\\
(citizen; New Zealand) & (citizen; U.S.) & (citizen; U.S.) & (child; $x$)& $\neg$(occup.; screenwriter), 1.0\\
(child; $y$) & (child; $z$) & (child; $u$) &  & $\neg \exists l $(instagram; $l$), 0.67\\
(occup.; screenwriter) & (occup.; screenwriter) & (occup.; screenwriter) & & $\neg w$(convicted; $w$), 0.33\\
(convicted; $v$)  & (instagram; $r$) & (instagram; $f$) & & $\neg$(citizen; New Zealand), 0.33\\
(instagram; $t$) & & & & \\
    \bottomrule
  \end{tabular}}
\end{table*}

\noindent
\textbf{Ranking Negative Statements.\ }
Given potentially large candidate sets, in a second step, ranking methods are needed. Our rationale in the design of the following four ranking metrics is to combine frequency signals with popularity and probabilistic likelihoods in a \emph{learning-to-rank model}.
\begin{enumerate}[noitemsep,topsep=0pt,parsep=0pt,partopsep=0pt]
\item \emph{Peer frequency (PEER):} The statement discovery procedure already provides a relative frequency, e.g., 0.9 of a given actor's peers are married, but only 0.1 are political activists. The former is an immediate candidate for ranking.
    \item \emph{Object popularity (POP):} When the discovered statement is of the form  $\neg$\term{(s; p; o)}, its relevance might be reflected by the popularity\footnote{Wikipedia page views.} of the \term{Object}. For example, $\neg$\term{(Brad Pitt; award; Oscar for Best Actor)} would get a higher score than $\neg$\term{(Brad Pitt; award; London Film Critics' Circle Award)}, because of the high popularity of the \textit{Academy Awards} over the \textit{London Film Awards}.
    \item \emph{Frequency of the Property (FRQ):} When the discovered statement has an empty \term{Object} \term{$\neg \exists x$(s; p; x)}, the frequency of the \term{Property} will reflect the authority of the statement. To compute the frequency of a \term{Property}, we refer to its frequency in the KB. For example, \term{$\neg \exists x$(Joel Slater; citizen; x)}  will get a higher score (4.1m citizenships in Wikidata) than \term{$\neg \exists x$(Joel Slater; twitter; x)} (294k Twitter usern\-ames).
    \item \emph{Pivoting likelihood (PIVO):} In addition to these \linebreak frequency/view-based metrics, we propose to consider textual background information about $e$ in order to better decide whether a negative statement is relevant. To this end, we build a set of statement pivoting classifier~\cite{razniewski2017doctoral}, i.e., classifiers that decide whether an entity has a certain statement (or property), each trained on the Wikipedia embeddings~\cite{wikipedia2vec} of 100 entities that have a certain statement (or property), and 100 that do not\footnote{On withheld data, linear regression classifiers achieve 74\% avg.\ accuracy on this task.}. To score a new statement (or property) candidate, we then use the pivoting score of the respective classifier, i.e., the likelihood of the classifier to assign the entity to the group of entities having that statement (or property).
\end{enumerate}

\noindent
The final score of a candidate statement is then computed as follows.

\begin{defn}[Ensemble Ranking Score]
\label{def:ensemble}
{\small
\begin{equation*}
Score=\begin{cases}
 \lambda_1 \text{PEER} + \lambda_2 \text{POP(\textit{o})} + \lambda_3 \text{PIVO}\\ \ \ if\ \ \ \neg(\textit{s; p; o}) \ is \ \mathit{satisfied}\\ \\
 \lambda_1 \text{PEER} + \lambda_4 \text{FRQ(\textit{p})} + \lambda_3 \text{PIVO}\\ \ \ if\ \ \ \neg \exists x (\textit{s; p; x}) \ is \ \mathit{satisfied}\\
\end{cases}
\end{equation*}}
\end{defn}

Hereby $\lambda_1$, $\lambda_2$, $\lambda_3$, and $\lambda_4$ are parameters to be tuned on data withheld from training.

%% file: orderedpeer.tex
\section{Order-oriented Peer-based Inference}
\label{sec:temporal}

In the previous section,  we assume a binary peer relation as the basis of peer group computation. In other words,  for each entity, any other entity is either a peer, or is not. Yet in expressive KBs, relatedness is typically graded and multifaceted, thus reducing this to a binary notion risks losing valuable information. We therefore investigate, in this section, how negative statements can be computed while using ordered peer set.

Orders on peers arise naturally when using real-valued similarity functions, such as Jaccard-similarity, or cosine distance of embedding vectors. An order also naturally arises when one uses temporal or spatial features for peering. Here are some examples:
\begin{enumerate}
    \item \emph{Spatial:} Considering the class \emph{national capital}, the peers closest to \textit{London} are \textit{Brussels} (199 miles), \textit{Paris} (213 miles), \textit{Amsterdam} (223 miles), etc.
    \item \emph{Temporal:} The same holds for temporal orders on attributes, e.g., via his role as president, the entities most related to \textit{Biden} are \textit{Trump} (predecessor), \textit{Obama} (pre-predecessor), \textit{Bush} (pre-pre-predecessor), etc.
\end{enumerate}

\noindent
\textbf{Formalization.\ }
Given a target entity $e_0$, a similarity function $\textit{sim}(e_a,e_b) \rightarrow \mathcal{R}$, and a set of candidate peers $E=\{e_1,...,e_n\}$, we can sort $E$ by $\textit{sim}$ to derive an ordered list of sets $L=[S_1, ..., S_n]$, where each $S_i$ is a subset of $E$ that consists of highly related entities to $e_0$.
\begin{example}
Let us consider temporal recency of having won the \textit{Oscars for Best Actor/Actress} as similarity function w.r.t. the target entity 
\emph{Olivia Colman}.   The ordered list of closest peer sets $S$ is 
\term{[\{Frances McDormand, Gary Oldman\}, \{Emma Stone, Casey Affleck\}, \{Brie Larson, L\-eonardo DiCaprio\}, \{Julianne Moore, Eddie Redmayne\}.., \{Janet Gaynor, Emil Jannings\}].}
\end{example}

Given an index of interest $m$ ($m \leq n)$, we have a prefix list $S_{[1,m]}$   of such an ordered peer set list $L$. For any negative statement candidate $\textit{stmt}$, we can     compute two ranking features:

\begin{enumerate}
    \item \emph{Prefix-volume (VOL)}: The prefix volume denotes the size of the prefix in terms of peer entities considered, i.e., $\textit{VOL}=|S_1 \cup ... \cup S_m|$. Note that the volume should not be mixed with the length $m$ of the prefix, which does not allow easy comparison, as sets may contain very different numbers of members.
    \item \emph{Peer frequency (PEER)}: As in Section~\ref{sec:inference}, \textit{PEER} denotes the fraction of entities in $S_1 \cup ... \cup S_m$ for which \textit{stmt} holds, i.e., $\textit{FRQ}$ / $\textit{VOL}$, where $\textit{FRQ}$ is the number of entities sharing the statement.
\end{enumerate}

Note that these two ranking features change values with prefix length. In addition, we can also consider static features like \textit{POP} and \textit{PIVO}, as introduced before.\\

Consider the entity $e$=\textit{Olivia Colman} from our example, with prefix length 3. For the statement \term{(citizen of; U.S.)}, $\textit{FRQ}$ is 5 and $\textit{VOL}$ is 6, i.e., unlike \emph{Olivia Colman}, 5 out of the 6 winners of the previous 3 years are U.S. citizens. Now considering prefix length 2, for the statement \term{(occupation;  director)}, $\textit{FRQ}$ is 1 and $\textit{VOL}$ is 4, i.e., unlike \emph{Olivia Colman}, 1 out of the 4 winners of the previous 2 years are directors.

We can now proceed to the actual problem of this section.

\medskip

\noindent
\textbf{Research Problem 2.\ }
Given an entity $e$ and an ordered set of peers, compile a ranked list of useful 
negative statements.\\


\noindent
\textbf{Ranking.\ }
What makes a negative statement from an ordered peer set \textit{informative}? It is easy to see that a statement is preferred over another, if it has both a higher peer frequency (\textit{PEER}) and prefix volume (\textit{VOL}). For example, the statement $\neg$\term{(citizen of; U.S.)} above is preferable over $\neg$\term{(occupation; director)}, due to it being both reported on a larger set of peers, and with higher relative frequency.  Yet statements can be incomparable along these two metrics, and this problem even arises when comparing a statement with itself over different prefixes: Is it more helpful if 3 out of the previous 4 winners are \textit{U.S.} citizens, or 7 out of the previous 10?

To resolve such situations, we propose to map the two features into a single one as follows:
\begin{equation}
\label{eqn:context}
score(\textit{stmt},L,m) = \lambda \cdot \textit{PEER} + (1-\lambda) \cdot log(\textit{FRQ})
\end{equation}

where $\lambda$ is again a parameter allowing to trade off the effects of the two variables. Note that we propose a logarithmic contribution of \textit{FRQ} - this is based on the rationale that larger number of peers is preferable. For example, for the same \textit{PEER} value 0.5, we can have a statement with 5 peers out of 10 and 1 peer out of 2.

\begin{figure}
 \caption{Retrieving useful negative statements about \textit{Olivia Colman}, using an ordered peer group.}
\includegraphics[width=\linewidth]{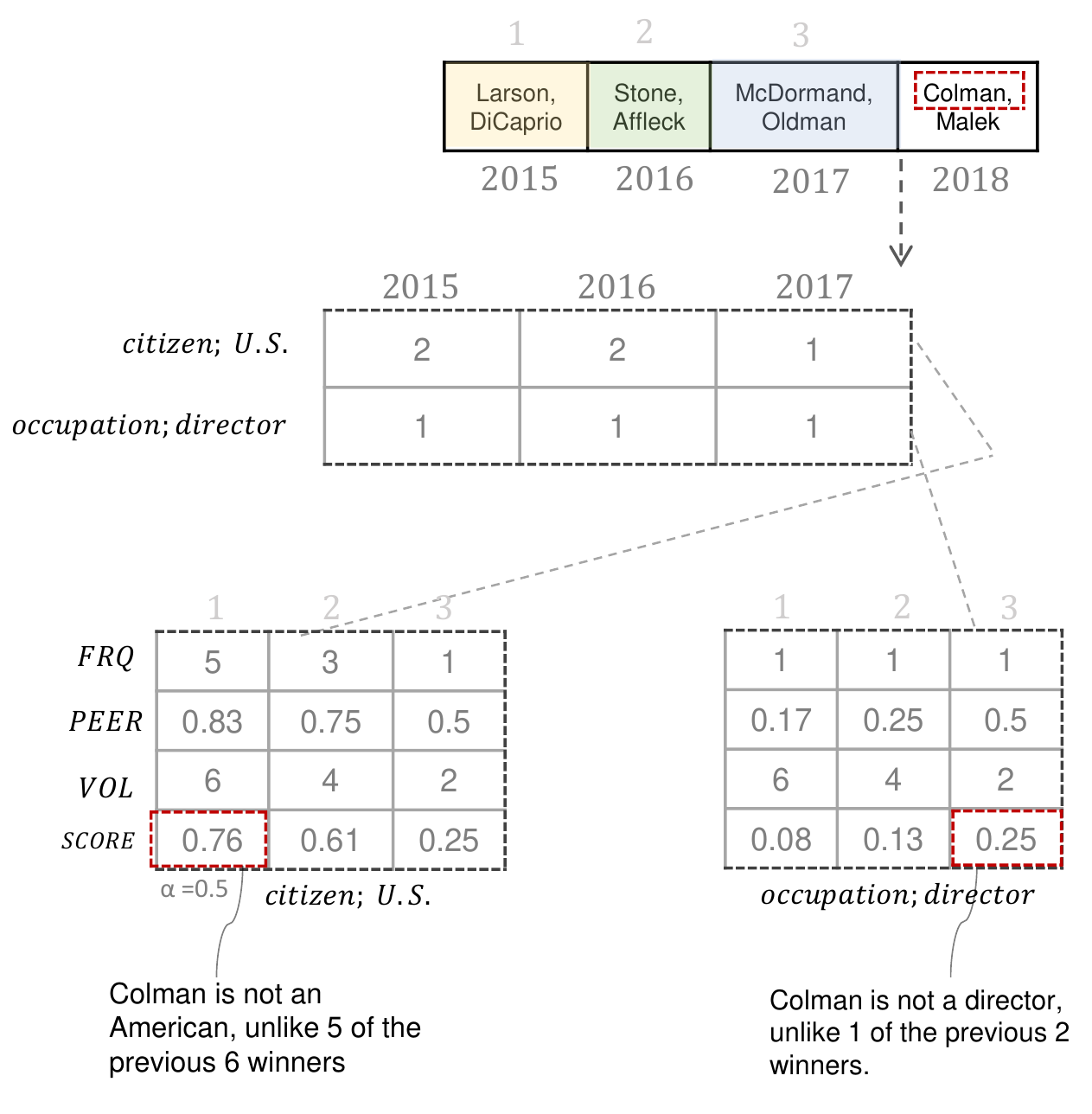}
\label{fig:diagram}
\end{figure}

Given the above example, the score for \emph{Olivia Colman}'s negative statement $\neg$\term{(citizen of; U.S.)} at prefix length 3 and $\alpha=0.5$ is $0.76$, with verbalization as ``unlike 5 of the previous 6 winners''. The same statement with prefix length 2 will receive a score of $0.61$, with verbalization as ``unlike 3 of the previous 4 winners''. As for $\neg$\term{(occupation; director)} at prefix length 3 and $\alpha=0.5$ is $0.08$, with verbalization as ``unlike 1 of the previous 6 winners''.  The same statement with prefix length 2 will receive a score of $0.13$, with verbalization as ``unlike 1 of the previous 4 winners''. This example is illustrated in Figure~\ref{fig:diagram}.\\

\noindent
\textbf{Computation.\ }
Having defined how statements over ordered peer sets can be ranked, we now present an efficient algorithm, Algorithm~\ref{alg:contextualizations}, to compute the optimal prefix length per statement candidate, based on a single pass over the prefix. Given the entity $e$=\textit{Olivia Colman}, ordered sets of her peers are collected in line 2.
\begin{equation*}
\begin{split}
L = [\text{winners of Oscar, winners of BAFTA,}\\ \text{..., recipients of CBE}].
\end{split}
\end{equation*}
For readability, we proceed with one ordered peer group, namely the winners of Oscar for Best Actor/Actress. The group contains ordered winners prior to $e$.
\begin{equation*}
\begin{split}
L_\text{winners of Oscar} = [\{\text{Frances McDormand, Gary Oldman}\},\\ \{\text{Emma Stone, Casey Affleck}\},\\ \{\text{Brie Larson, Leonardo DiCaprio}\},\\ \{\text{Julianne Moore, Eddie Redmayne}\}\\ \text{..,}\\ \{\text{Janet Gaynor, Emil Jannings}\}].
\end{split}
\end{equation*}
Similar to the previous algorithm, all statements of the peers are then retrieved from the KB (line 11 and 12). For every candidate statement $st$, the score(s) of the statement is computed with different prefix lengths (loop at line 27), starting with $pos$ (position of $e$ in the ordered set) and stopping at the start position 1. The maximum score is then returned with its corresponding values of \textit{FRQ} and \textit{VOL}, i.e., $max\_\mathit{frq}$ and $max\_\mathit{vol}$ (line 37). The returned candidate statement with its highest score (within one ordered group of peers $L_i$) is compared across many ordered groups of peers (i.e., other groups in $L$), to be either replaced or disregarded from the final list of negations $N$.

\begin{algorithm*}[t!]
    \SetKwInOut{Input}{Input}
    \SetKwInOut{Output}{Output}
    \Input{\small knowledge base $\mathit{KB}$, entity $e$, ordered peer collection function \textit{ordered\_peers}, \small number of results $k$, \small hypeparameter of scoring function $\alpha$}
    \Output{ \small top-$k$ negative statement candidates for $e$}
        
        \textit{\textbf{L}}$[]$= \textit{ordered\_peers}$(e)$ \Comment{List of ordered peer group(s); Group $L_i$ at position $i$ is one ordered group (list).}\\
         \textit{\textbf{N}}$[]$= $\emptyset$ \Comment{\small Ranked list of negative statements about $e$.}\\
    \For{$L_i$ $\in$ \textbf{L}}{
        $candidates$ = [] \\
        $pos$=position($L_i$, $e$) \Comment{\small Position of $e$ in the ordered set.}\\
        \For{$pe$ $\in$ $L_i$}{ 
            \If{$pe$ == $p$}{continue}
            $candidates$+=$collectP(pe)$\\
            $candidates$+=$collectPO(pe)$
        }
        $ucandidates$ = $unique(candidates)$\\
        \For{$st$ $\in$ $ucandidates$}{
            $sc$ = $scoring(st, L_i, e, pos, \alpha)$ \Comment{\small Dynamic scoring of every statement st with different prefix lengths.}\\
            \If{$getnegation(N, st).score$ $<$ $sc$}{ $setscore(N, st, sc)$}
        }
     }
        $N$-=$\mathit{inKB}(e,N)$\\
        return $max(N,k)$
    \texttt{\\}
    \texttt{\\}
    \SetKwFunction{FMain}{scoring}
    \SetKwProg{Fn}{Function}{:}{}
    \Fn{\FMain{$st$, $S$, $e$, $pos$, $\alpha$}}{
        max\_sc = - $\inf$; max\_frq = - $\inf$; max\_vol = - $\inf$; \Comment{Initializing the maximum score, frequency, and volume for statement $st$.}\\
        frq = 0; vol=0; \Comment{Initializing the frequency and volume of statement $st$.}\\
         \For{$j$ = $pos$; $j$ $>=$ 1; $j{-}{-}$}
             {
                vol += countentities($S[j]$) \Comment{\small Computing number of entities at position $j$.}\\
                frq += countif($st$, $candidates$, $S[j]$) \Comment{\small Computing number of entities at position $j$ that share $st$.}\\
                sc = $\alpha * \frac{frq}{vol} + (1-\alpha)* log(frq)$ \Comment{Computing the score of $st$ at position $j$.}\\
                 \If{sc $>$ max\_sc}{
                 max\_sc = sc;\\
                 max\_frq = frq;\\
                 max\_vol = vol\; 
                }
         }}
         \textbf{return} max\_sc, max\_frq, max\_vol\\
\caption{Order-oriented peer-based candidate retrieval algorithm.}
\label{alg:contextualizations}
\end{algorithm*}

%% file: conditional.tex
\section{Conditional Negative Statements}
\label{sec:restricted}

In our negation inference methods, we generate two classes of negative statements, grounded negative statements, and universally negative statements. These two classes represent extreme cases: each grounded statement negates just a single assertion, while each universally negative statement negates all possible assertions for a property. Consequently, grounded statements may make it difficult to be concise, while universally negative statements do not apply whenever at least one positive statement exists for a property.
A compromise between these extremes is to restrict the scope of universal negation. For example, it is cumbersome to list all major universities that \textit{Einstein} did not study at, and it is not true that he did not study at any university. However, salient statements are that he \textit{did not study at any U.S. university}, or that he \textit{did not study at any private university}.
We call these statements \emph{conditional negative statements}, as they represent a conditional case of universal negation. In principle, the conditions used to constrain the object could take the form of arbitrary logical formulas. For proof of concept, we focus here on conditions that take the form of a single triple pattern.

\begin{defn}
A conditional negative statement takes the form $\neg \exists o$: (s; p; $o$), (o; p'; o'). It is satisfied if there exists no $o$ such that (s; p; $o$) and  (o; p'; o') are in $K^i$.
\end{defn}


In the following, we call the property $p'$ the \textit{aspect} of the conditional negative statement.

\begin{example}
Consider the statement that Einstein did not study at any \textit{U.S.} university. It could be written as $\neg\exists o:$ \term{(Einstein; education; o)}, \term{(o; located in; U.S.)}. It is true, as \textit{Einstein} only studied at \textit{ETH Zurich, Luitpold-Gymnasium, Alte Kantonsschule Aarau}, and \textit{University of Zurich}, located in \textit{Switzerland} and \textit{Germany}. Another possible conditional negative statement is $\neg\exists o:$ \term{(Einstein; education; o)}, \term{(o; type; private University)}, as none of these schools are private.  
\end{example}

As before, the challenge is that there is a near-infinite set of true conditional negative statements, so a way to identify interesting ones is needed. For example, \textit{Einstein} also did not study at any \textit{Jamaican} university, nor did he study at any university that \textit{Richard Feynman} studied at, etc. 
One way to proceed would be to traverse the space of possible conditional negative statements, and score them with another set of metrics. Yet compared to universally negative statements, the search space is considerably larger, as for every property, there is a large set of possible conditions via novel properties and constants (e.g., \textit{``that was located in Armenia/Brazil/China/Denmark/...''}, \textit{``that was attended by Abraham/Beethoven/Cleopatra/...''}). 
So instead, for efficiency, we propose to make use of previously generated grounded negative statements: In a nutshell, the idea is first to generate grounded negative statements, then in a second step, to \emph{lift} subsets of these into more expressive conditional negative statements. 
A crucial step is to define this lifting operation, and what the search space for this operation is.

With the \textit{Einstein} example, shown in Table~\ref{tab:einsteinlifting}, we could start from three relevant grounded negative statements that \textit{Einstein} did not study at \textit{MIT, Stanford}, and \textit{Harvard}. One option is to lift them based on aspects they all share: their locations, their types, or their memberships. The values for these aspects are then automatically retrieved: they are all located in the \textit{U.S.}, they are all private universities, they are all members of the \textit{Digital Library Federation}, etc., however, not all of these may be interesting. So instead we propose to \emph{pre-define} possible aspects for lifting, either using manual definition, or using methods for facet discovery, e.g., for faceted interfaces~\cite{oren2006extending}. For manual definition, we assume the condition to be in the form of a single triple pattern. A few samples are shown in Table~\ref{tab:aspects}. For \term{educated at}, it would result in statements like ``e was not educated in the \textit{U.K.}'' or ``e was not educated at a public university''; for \term{award received}, like ``e did not win any category of \textit{Nobel Prize}''; and for \term{position held}, like ``e did not hold any position in the \textit{House of Representatives}''.\\

\noindent
\textbf{Research Problem 3.\ }
Given a set of grounded negative statements about an entity $e$, compile a ranked list of useful conditional negative statements.\\

We propose an approach with Algorithm~\ref{alg:restricted}. Consider $e$=\emph{Einstein}, and the set of possible aspects $\mathit{ASP}$ for lifting containing only two aspects about \term{educated at}, for readability. 
\begin{equation*}
ASP = [(\text{educated at: located in, instance of})].
\end{equation*}
The three grounded negative statements about \emph{Einstein} with \term{educated at} property are:
\begin{equation*}
\mathit{NEG}=[\neg(\text{educated at: MIT, Stanford, Harvard})].
\end{equation*}
The loop at line 2 considers every property ($neg.p$) in $\mathit{NEG}$ (e.g., \term{educated at}), and collect its aspects at line 3. For this example, the list of aspects $asp$ for this predicate consists of the location and the type of the educational institution.
\begin{equation*}
asp=[\text{located in, instance of}].
\end{equation*}
At line 4, the loop visits every aspect $a$ in $asp$ and look for aspect values (i.e., the locations and types of Einstein's schools). $neg.o$ are the objects that share the same predicate in the grounded negative statements list.
\begin{equation*}
neg.o=[\text{MIT, Stanford, Harvard}].
\end{equation*}
For every object $o$, aspect values are collected and their relative frequencies are stored. For readability, line 6 is only a high level version of this step. As mentioned before, the aspects are manually pre-defined and their values are automatically retrieved.
\begin{equation*}
 \begin{aligned}
getaspvalues(\text{Wikidata, located in, MIT}) &= [\text{U.S}].\\ 
getaspvalues(\text{Wikidata, located in, Stanford}) &= [U.S].\\
getaspvalues(\text{Wikidata, located in, Harvard}) &= [U.S].\\
getaspvalues(\text{Wikidata, instance of, MIT}) &=\\ [\text{institute of technology, private university}].\\
getaspvalues(\text{Wikidata, instance of, Stanford}) &=\\ [\text{research university, private university}].\\
getaspvalues(\text{Wikidata, instance of, Harvard}) &=\\ [\text{research university, private university}].\\
 \end{aligned}
\end{equation*}
Hence the aspect value for \term{educated at}, namely \term{(located in; U.S.)} receives a score of 3, and is added to the conditional negation list $\mathit{cond\_NEG}$. After retrieving and scoring all the aspect values, the top-2 (with $k$ =2) conditional negative statements are returned. In this example, the final results are $\mathit{cond\_NEG}$ = \term{[($\neg\exists o$(Einstein; educated at; $o$) ($o$; located in; U.S.), 3)}, (\term{$\neg\exists o$(Einstein; educa\-ted at; $o$) ($o$, instance of; private university), 3)}].

\begin{table*}
  \caption{Negative statements about \textit{Einstein}, before and after lifting.}
  \label{tab:einsteinlifting}
  \centering
  \scalebox{0.9}{
   \begin{tabular}{l|l}
    \toprule
    \multicolumn{1}{c}{\textbf{Grounded negative statements}} & \multicolumn{1}{c}{\textbf{Conditional negative statements}}\\
    \midrule
$\neg$(educated at; MIT) & $\neg\exists o$(educated at; $o$) ($o$; located in; U.S.)\\
$\neg$(educated at; Stanford) & $\neg\exists o$(educated at; $o$) ($o$, instance of; private university)\\
$\neg$(educated at; Harvard) & \\
    \bottomrule
  \end{tabular}
  }
  \end{table*}
  
\begin{algorithm*}[t!]
    \SetKwInOut{Input}{Input}
    \SetKwInOut{Output}{Output}
    \Input{\small knowledge base $\mathit{KB}$, entity $e$, aspects $\mathit{ASP}$ = [($x_{1}$: $y_{1}$, $y_{2}$, ..), ..., ($x_{n}$: $y_{1}$, $y_{2}$, ..)], \small grounded negative statements about $e$ $\mathit{NEG}$ = [$\neg$($p_{1}$: $o_{1}$, $o_{2}$, ..), ..., $\neg$($p_{m}$: $o_{1}$, $o_{2}$, ..)], \small number of results $k$}
    \Output{\small $k$-most frequent conditional negative statements for $e$}
    \textbf{$\mathit{cond\_NEG}$}= $\emptyset$ \Comment{\small Ranked list of conditional negations about $e$.}\\
           \For{$neg.p$ $\in$ $\mathit{NEG}$}{ 
           $\mathit{asp}$ = $getspects(neg.p, ASP)$ \Comment{Retrieving aspects of predicate $neg.p$.}\\
           \For{$a$ $\in$ $asp$}{
           \For{$o$ $\in$ $neg.o$}{ 
            $\mathit{cond\_NEG}$ += getaspvalues($\mathit{KB}$, $a$, $o$) \Comment{Collecting aspect values about $o$.}
           }
           }
        }
        $\mathit{cond\_NEG}$-=$inKB(e,\mathit{cond\_NEG})$\\
        return $max(\mathit{cond\_NEG},k)$
\caption{Lifting grounded negative statements algorithm.}
\label{alg:restricted}
\end{algorithm*}

\begin{table}
  \caption{A few samples of property aspects.}
  \label{tab:aspects}
  \centering
  \scalebox{0.8}{
   \begin{tabular}{l|l}
    \toprule
    \multicolumn{1}{c}{\textbf{Property}} & \multicolumn{1}{c}{\textbf{Aspect(s)}}\\
    \midrule
educated at & located in; instance of;\\
award received & subclass of;\\
position held & part of;\\
    \bottomrule
  \end{tabular}
  }
  \end{table}

%% file: experiments.tex
\section{Experimental Evaluation}
\label{sec:experiments}

\subsection{Peer-based Inference}
\label{subsec:similarityexp}
\noindent
\textbf{Setup.\ }We instantiated the peer-based inference method with 30 peers, popularity based on Wikipedia page views, and peer groups based on entity occupations.
The choice of this simple peering function was inspired by Recoin~\cite{RECOIN}.
In order to further ensure relevant peering, we also only considered entities as candidates for peers, if their Wikipedia viewcount was at least a quarter of that of the subject entity.
We randomly sampled 100 popular Wikidata people. For each of them, we collected 20 negative statement candidates: 10 with the highest \textit{PEER} score, 10 being chosen at random from the rest of retrieved candidates. We then used crowdsourcing\footnote{\url{https://www.mturk.com}} to annotate each of these 2000 statements on whether it was interesting enough to be added to a biographic summary text (Yes/Maybe/No). Each task was given to 3 annotators. Interpreting the answers as numeric scores (1/0.5/0), we found a standard deviation of 0.29, and full agreement of the 3 annotators on 25\% of the questions. Our final labels are the numeric averages among the 3 annotations.

\noindent
\textbf{Parameter Tuning.\ } To learn optimal parameters for the ensemble ranking function (Definition~\ref{def:ensemble}), we trained a linear regression model using 5-fold cross validation on the 2k labels for usefulness. Four example rows are shown in Table~\ref{tab:training}. Note that the ranking metrics were normalized using a ranked transformation to obtain a uniform distribution for every feature.

The average obtained optimal parameter values were -0.03 for \textit{PEER}, 0.09 for \textit{FRQ(p)}, -0.04 for \textit{POP(o)}, and 0.13 for \textit{PIVO},  and a constant value of 0.3., with a 71\% out-of-sample precision.

\begin{table*}
  \caption{Data samples for illustrating parameter tuning.}
  \label{tab:training}
 \resizebox{\textwidth}{!}{\begin{tabular}{llllll}
    \toprule
    \multicolumn{1}{c}{\bf{Statement}} & \multicolumn{1}{c}{\bf{PEER}} & \multicolumn{1}{c}{\bf{FRQ(p)}}& \multicolumn{1}{c}{\bf{POP(o)}}& \multicolumn{1}{c}{\bf{PIVO}}&\multicolumn{1}{c}{\bf{Label}}\\
    \midrule
$\neg$(Bruce Springsteen; award; Grammy Lifetime Achievement Award) & 0.8 & 0.8 & 0.55 & 0.25 & 0.83\\
$\neg$(Gordon Ramsay; lifestyle; mysticism) & 0.3 & 0.8 & 0.8 & 0.65 & 0.33\\
$\neg \exists x$(Albert Einstein; doctoral student; x) & 0.85 & 0.9 & 0.15 & 0.4 & 0.66\\
$\neg \exists x$(Celine Dion; educated at; x) & 0.95 & 0.95 & 0.25 & 0.95 & 0.5\\
    \bottomrule
  \end{tabular}}
  \end{table*}
\noindent
\textbf{Ranking Metric.\ }To compute the ranking quality of our method against a number of baselines, we used the Discounted Cumulative Gain (DCG)~\cite{NDCG}, which is a measure that takes into consideration the rank of relevant statements and can incorporate different relevance levels. DCG is defined as follows:
\begin{equation*}
DCG(i)=\begin{cases}
& G(1)\ \ \ if\ $i=1$\\
 & DCG(i-1)+\frac{G(i)}{log(i)} \ \ \ \text{\textit{otherwise}}
\end{cases}
\end{equation*}
where i is the rank of the result within the result set, and $G(i)$ is the relevance level of the result. We set $G(i)$ to a value between 1 and 3, depending on the annotator's assessment. We then averaged, for each result (statement), the ratings given by all annotators and used it as the relevance level for the result. Dividing the obtained DCG by the DCG of the ideal ranking, we obtained the normalized DCG (nDCG), which accounts for the variance in performance among queries (entities).

\noindent
\textbf{Baselines.\ }We used three \textit{baselines}: As a naive baseline, we randomly ordered the 20 statements per entity. This baseline gives a lower bound on what any ranking model should exceed. We also used two competitive embedding-based baselines, TransE~\cite{transE} and HolE~\cite{holE}. For these two, we used pretrained models, from~\cite{ho2018rule}, on Wikidata (300k statements) containing prominent entities of different types, which we enriched with all the statements about the sampled entities. We plugged their prediction score for each candidate grounded negative statement.\footnote{Note that both models are not able to score statements about universal absence, a trait shared with the object popularity heuristic in our ensemble.}

\begin{table*}
  \caption{Ranking metrics evaluation results for peer-based inference.}
  \label{tab:rankingNDCG}
  \centering
  \resizebox{0.8\textwidth}{!}{\begin{tabular}{llllll}
    \toprule
    \multicolumn{1}{l}{\bf{Ranking Model}} &\multicolumn{1}{c}{\bf{Coverage(\%)}} & \multicolumn{1}{c}{$\boldsymbol{nDCG_3}$} & \multicolumn{1}{c}{$\boldsymbol{nDCG_5}$}& \multicolumn{1}{c}{$\boldsymbol{nDCG_{10}}$}& \multicolumn{1}{c}{$\boldsymbol{nDCG_{20}}$}\\
    \midrule
Random & 100 & 0.37 & 0.41 & 0.50 & 0.73\\
TransE~\cite{transE} & 31 & 0.43 & 0.47 & 0.55 & 0.76\\
HolE~\cite{holE} & 12 &	0.44 & 0.48 & 0.57 & 0.76\\
\midrule
Property Frequency & 11 & \bf{0.61} & \bf{0.61} & \bf{0.66} & \bf{0.82}\\
Object Popularity & 89 & 0.39 & 0.43 & 0.52 & 0.74\\
Pivoting Score & 78 & 0.41 & 0.45 & 0.54 & 0.75\\
Peer Frequency & 100 &	\bf{0.54} &	\bf{0.57} &	\bf{0.63} &	\bf{0.80}\\
\midrule
Ensemble & 100 &	\bf{0.60} &	\bf{0.61} &	\bf{0.67} & \bf{0.82}\\
    \bottomrule
  \end{tabular}}
  \end{table*}
  \noindent
  \textbf{Results.\ }Table \ref{tab:rankingNDCG} shows the average $nDCG$ over the 100 entities for top-k negative statements for k equals 3, 5, 10, and 20. As one can see, our ensemble outperforms the best baseline by 6 to 16\% in $nDCG$. The coverage column reflects the percentage of statements that this model was able to score. For example, for the \textit{Popularity of Object}, $POP(o)$ metric, a universally negative statement will not be scored. The same applies to TransE and HolE.
  
 Ranking with the \textit{Ensemble} and ranking using the \textit{Frequency of Property} outperforms all other ranking metrics and the three baselines, with an improvement over the random baseline of 20\% for k=3 and k=5. Examples of ranked top-3 negative statements for \textit{Albert Einstein} are shown in Table \ref{tab:rank_qualitative}. The random rank basically display any candidate negation if it holds for at least one peer. For instance, \textit{Omar Sharif} is \textit{Einstein}'s peer under the \textit{non-fiction writer} group. This makes the negation ``Tarek Sharif not a child of Einstein'' possible, hence, the necessity for a ranking step. Moreover, \textit{Omar Sharif} is also an actor, which brings other topics to the result set of \textit{Einstein}, such as not winning \textit{film awards}. This is where peer frequency makes a difference, i.e., most of \textit{Einstein}'s peers are \textit{not} actors. By relying on the property frequency for ranking, we can see that only universally absent statements get the highest scores. Even though it displays interesting negations (e.g., despite his status as famous researcher, \textit{Einstein} truly never formally supervised any PhD student), the top-k result set lacks grounded negative statements. Ensemble ranking, on the other hand, takes into consideration several features simultaneously, and covers both classes of negation. It returns interesting statements such as that \textit{Einstein} notably refused to work on the \textit{Manhattan} project, and was suspected of communist sympathies.

\begin{table*}
  \caption{Top-3 results for \textit{Albert Einstein} using 3 ranking metrics.}
  \label{tab:rank_qualitative}
  \centering 
   \resizebox{0.8\textwidth}{!}{\begin{tabular}{l|l|l}
    \toprule
    \multicolumn{1}{l}{\textbf{Random rank}} &    \multicolumn{1}{l}{\textbf{Property frequency}} & \multicolumn{1}{l}{\textbf{Ensemble}}\\
\toprule
\multicolumn{1}{l}{$\neg \exists x$(instagram; x)} &    \multicolumn{1}{l}{$\neg \exists x$(doctoral student; x)} & \multicolumn{1}{l}{$\neg$(occup.; astrophysicist)}\\
 \multicolumn{1}{l}{$\neg$(child; Tarek Sharif)} & \multicolumn{1}{l}{$\neg \exists x$(candidacy in election; x)} & \multicolumn{1}{l}{$\neg$(party; Communist Party USA)}\\
  \multicolumn{1}{l}{$\neg$(award; BAFTA)} & \multicolumn{1}{l}{$\neg \exists x$(noble title; x)} & \multicolumn{1}{l}{$\neg \exists x$(doctoral student; x)}\\
    \bottomrule
  \end{tabular}}
  \label{tbl:may:einstein}
\end{table*}

\noindent
\textbf{Correctness Evaluation.\ } We used crowdsourcing to assess the correctness of results from the peer-based method. We collected 1k negative statements belonging to the three types, namely people, literature work, and organizations. Every statement was annotated 3 times as either correct, incorrect, or ambiguous. 63\% of the statements were found to be correct, 31\% were incorrect, and 6\% were ambiguous. Most incorrect statements are due to KB completion issues. Interpreting the scores numerically (0/0.5/1), annotations showed a standard deviation of 0.23. 

\noindent
\textbf{PCA (Partial Completeness Assumption) vs.\ CWA\ }
For a sample of 200 statements about people (10 each for 20 entities), half generated only relying on the CWA, half additionally filtered to satisfy the PCA (subject has at least one other object for that property~\cite{AMIEP}), we manually checked correctness.
We observed  84\% accuracy
for PCA-based statements, and 57\% for CWA-based statements. So the PCA yields significantly more correct negative statements, though losing the ability to predict universally negative statements.

\noindent
\textbf{Subject coverage.} Our peer-based inference method offers a very high subject coverage and is able to discover negative statements about almost any existing entity in a given KB, whereas for pre-trained embedding-based baselines, many subjects are out-of-vocabulary, or come with too little information to predict statements.


\subsection{Inference with Ordered Peers}
\label{sub:temporalexperiments}

In the following, we used temporal order on specific roles, or on specific attribute values, to compute ordered peer sets. In particular, we used two common forms of temporal information in Wikidata to compute such peer groups: 
\begin{itemize}
    \item \textbf{Time-based Qualifiers (TQ)}: Temporal qualifiers are time signals associated with statements about entities. In Wikidata, some of those qualifiers are \textit{point in time} (P585), \textit{start time} (P580), and \textit{end time} (P582). A few samples are shown in Table~\ref{tab:qualifiers}.
    \item \textbf{Time-based Properties (TP)}: Temporal properties are properties like \textit{follows} (P155) and \textit{followed by} (P156) indicating a chain of entities, ordered from oldest to newest, or from newest to oldest. For instance, \term{[The Cossacks; followed by; War and Peace; followed by; Anna Karenina; ..]}\footnote{Novels of by Leo Tolstoy.}
\end{itemize}

We created TQ groups from aggregating information about people sharing the same statements. For example, \term{position held; President of the U.S.} is one TQ group, where members will have a \textit{start time} for this position, as well as an \textit{end time}. In case of absence of an \textit{end time}, this implies that the statement holds to this day (\term{Donald Trump}'s statement in Table~\ref{tab:qualifiers}). In other words, we aggregated entities sharing the same predicate-object pair, which will be treated as the peer group's title, and ranked them in ascending order of time qualifiers. For the \textit{point in time} qualifier, we simply ranked the dates from oldest to newest, and for the \textit{start/end date}, we ranked the end date from oldest to newest.
If the \textit{end date} is missing, the entity will be moved to the newest slot.

We collected a total of 19.6k TQ groups (13.6k using the \textit{start/end date} qualifier and 6k using the \textit{point in time} qualifier). Based on a manual analysis of a random sample of 100 groups of different sizes, we only considered time series with at least 10 entities\footnote{This variable can be easily adjusted depending on the preference of the developers and/or the purpose of the application.}.

We created TP groups by first collecting all entities reachable by one of the transitive properties, \textit{follows} (P155) and \textit{followed by} (P156). Considering each of the collected entities as a source entity, we computed the longest possible path of entities with only transitive properties. This path consists in an ordered set of peers. To avoid the problem of double-branching (one entity followed by two entities), we considered the two directions separately. Again, one path will be chosen at the end; the one with maximum length. 
The total number of TP groups is 19.7k groups. We limited the size of the groups to at least 10 and at most 150\footnote{We did not truncate the groups, we simply disregarded any group smaller or larger than the thresholds.}. 

\begin{table*}
  \caption{Samples of temporal information in Wikidata.}
  \label{tab:qualifiers}
  \centering
  \scalebox{0.9}{
   \begin{tabular}{l|l}
    \toprule
    \multicolumn{1}{c}{\textbf{Statement}} & \multicolumn{1}{c}{\textbf{Time-based qualifier(s)}}\\
    \midrule
(Barack Obama; position held; U.S. senator) & {\small \textit{start time}}: 3 January 2005; {\small \textit{end time}}: 16 November 2008\\
(Maya Angelou; award received; Presidential Medal of Freedom) & {\small \textit{point in time}}: 2010\\
(Donald Trump; spouse; Melania Trump) & {\small \textit{start time}}: 22 January 2005\\
    \bottomrule
  \end{tabular}
  }
  \end{table*}
  
\noindent
\textbf{Setup and Baseline.} We chose 100 entities, that belongs to at least one ordered set of peers, from Wikidata: 50 people and 50 literature works. We collected top-5 negative statements for each of those entities (for people, we consider TQ groups, and for literature works, TP groups). We made this choice because of the lack of entities of type person with transitive properties. In case an entity belongs to several groups, we merged all the results it is receiving from different groups, ranked them, and retrieved the top-5 statements. Similarly, as a baseline, using the peer-based inference method of Section~\ref{sec:inference}, instantiated with cosine similarity on Wikipedia embeddings~\cite{wikipedia2vec} as similarity function, we collected the top-5 negative statements for the same entities. We ended up with 1k statements, 500 inferred by each model.

\noindent
\textbf{Correctness Evaluation.} We randomly retrieved 400 negative statements from the 1k statements collected above, 200 from each model (100 about people, and 100 about literature works). We then assessed the correctness of each method using crowdsourcing. We showed each statement to 3 annotators, asking them to choose whether this statement is correct, incorrect, or ambiguous. Results are shown in Table~\ref{tab:simvstempcorrec}. Our order-oriented inference method clearly infers less incorrect statements by 9 percentage points for people, and 5 for literature works. It also produces more correct statements for people by 10 percentage points, and literature work by 3. The percentage of queries with full agreement in this task is 37\%. Also, annotations show a standard deviation of 0.17.
\begin{table}
\centering
  \caption{Correctness of order-oriented and peer-based methods.}
  \label{tab:simvstempcorrec}
  \begin{tabular}{llc}
    \toprule
   & \multicolumn{1}{l}{\bf{People}}  & \multicolumn{1}{c}{\bf{Literature Work}}  \\
    \midrule
    \multicolumn{3}{c}{Peer-based inference}\\
    \midrule
           & \multicolumn{1}{l}{\bf{\%}}  & \multicolumn{1}{c}{\bf{\%}}  \\
    \midrule
Correct & 81 & 88\\
Incorrect &  18 & 12\\
Ambiguous & 1 & 0\\
\midrule
\multicolumn{3}{c}{Order-oriented inference}\\
\midrule
       & \multicolumn{1}{l}{\bf{\%}}  & \multicolumn{1}{c}{\bf{\%}}  \\
    \midrule
Correct & \bf{91} & \bf{91}\\
Incorrect & 9 & 7\\
Ambiguous & 0 & 2 \\
    \bottomrule
  \end{tabular}
  
  \end{table}

\noindent
\textbf{Subject Coverage.} To assess the subject coverage of the order-oriented method, we randomly sampled 1k entities from each dataset, and tested whether it is a member of at least one ordered set, thus the ability to infer useful negative statements about it. For TQ groups, we randomly sampled 1k people, which results in a  coverage of 54\%. And for TP groups, we randomly sampled 1k literature works, and also received a coverage of 54\%. Although the order-oriented method produces better negative statements on both notions of correctness and usefulness (as we will see next), it does not outperform the baseline on subject coverage. However, using a different function to order peers might affect this drastically (e.g., using real-valued similarity functions like cosine distance of embeddings).

\noindent
\textbf{Usefulness.} To assess the quality of our inferred statements from the order-oriented inference method against the baseline (the peer-based inference method), we presented to the annotators two sets of top-5 negative statements about a given entity, and asked them to choose the more interesting set. The total number of opinions collected, given 100 entities, 3 annotations each, is 300. To avoid biases, we repeatedly switched the position of the sets. Results are shown in Table~\ref{tab:WDinterestingness}. Overall results show that our method is preferred by 10\% of the entities for both domains. The standard deviation of this task is 0.24 and the percentage of queries with full agreement is 18\%. We observe two advantages of the ordered set of peers over the previous method: i) it gives better interpretations of what a peer is, by automatically producing labels for peer groups (e.g., Presidents of the U.S., Winners of the Best Actor Academy Award); and ii) it maximizes the \textit{peerness} within a group. For instance, with Wikipedia embedding~\cite{wikipedia2vec}, closest peers to \textit{Donald Trump} are \textit{Hillary Clinton} and \textit{Donald Trump Jr.}. While the peerness with the input entity is obvious, there is not much similarity between the peers themselves, hence, very sparse candidate negations. However, with the order-oriented peering, \textit{Trump}'s peers include \textit{Barack Obama} and \textit{George W. Bush}, who are also peers of each other. 

\noindent
\textbf{Evaluation of Verbalizations.} One main contribution that our order-oriented inference method offers are \textit{verbalizations} produced with every inferred negative statement. In other words, it can, unlike the peer-based inference method, produce more concrete explanations of the usefulness of the inferred negations. For example, the inferred negative statement \term{$\neg$(Abraham Lincoln; cause of death; natural} \term{causes)} was inferred by both of our methods. However, each method offers a different verbalization. For the peer-based method, the verbalization is ``unlike 10 of 30 similar people'', and for the order-oriented method is ``unlike 12 of the previous 12 presidents of the U.S.''. To assess the quality of the verbalizations more formally, we conducted a crowdsourcing task with 100 useful negations that were inferred by both methods from our previous experiment. For every negative statement, the annotator was shown two different verbalizations on ``why is this negative statement noteworthy''. We asked the annotator to choose the better verbalization, she can choose Verbalization1, Verbalization2, or Either/Neither. Results show that verbalizations produced by our order-oriented inference method were chosen 76\% of the time, by the peer-based inference method 23\% of the time, and the either or neither option only 1\% of the time. The standard deviation is 0.23, and the percentage of queries with full agreement is 20\%. Table~\ref{tab:explanations} shows a number of examples, using different grouping functions for the peer-based method.

\begin{table*}
  \caption{Negative statements and their verbalizations using peer-based and order-oriented methods.}
  \label{tab:explanations}
 \resizebox{\textwidth}{!}{\begin{tabular}{llll}
    \toprule
    \multicolumn{1}{c}{\bf{Statement}} & \multicolumn{1}{c}{\bf{Order-oriented}} & \multicolumn{1}{c}{\bf{Peer-based}}& \multicolumn{1}{c}{\bf{Peering}}\\
      \multicolumn{1}{c}{\bf{}} & \multicolumn{1}{c}{\textit{Unlike..}} & \multicolumn{1}{c}{\textit{Unlike..}}& \multicolumn{1}{c}{}\\
    \midrule
   $\neg$(Emmanuel Macron; member;  National Assembly)	& 29 of 36 members of La République En Marche party & 70 of 100 similar people &	WP embed.~\cite{wikipedia2vec}\\
   $\neg$(Tim Berners-Lee; citizenship; U.S.) & 101 of previous 115 winners of the MacArthur Fellowship & 53 of 100 sim. comp. scientists &	Structured facets\\
 $\neg$(Michael Jordan; occupation; basketball coach) & 27 of prev. 49 winners of the NBA All-Defensive Team & 31 of 100 sim. people & WP embed.~\cite{wikipedia2vec}\\
 $\neg$(Theresa May; position; Opposition Leader) & 11 of prev. 14 Leaders of the Conservative Party & 10 of 100 sim. people &	WP embed.~\cite{wikipedia2vec}\\
 $\neg$(Cristiano Ronaldo; citizenship; Brazil) & 4 of prev. 7 winners of the Ballon d'Or & 20 of 100 sim. football players & Structured facets\\
    \bottomrule
  \end{tabular}}
  \end{table*}

\begin{table}
\centering
  \caption{Usefulness of order-oriented and peer-based methods.}
  \label{tab:WDinterestingness}
  \begin{tabular}{llc}
    \toprule
    & \multicolumn{1}{l}{\bf{People}}  & \multicolumn{1}{c}{\bf{Literature Work}}  \\
    \midrule
           &     \multicolumn{1}{l}{\bf{\%}}  & \multicolumn{1}{c}{\bf{\%}}\\
           \midrule
    Peer-based inference & 42 & 44\\
Order-oriented inference  & \bf{52} & \bf{54} \\
Both & 6 & 2\\
    \bottomrule
  \end{tabular}
  
  \end{table}

\subsection{Conditional Negative Statements Evaluation}
\label{subsec:restrictedexp}

We evaluated our lifting technique to retrieve useful conditional negative statements, based on three criteria: (i) compression, (ii) correctness, and (iii) usefulness. We collected the top-200 negative statements about 100 entities (people, organizations, and art work), and then applied lifting on them.

\noindent
\textbf{Compression.}   On average, 200 statements are reduced to 33, which means that lifting compresses the result set by a factor of 6.

\noindent
\textbf{Correctness.} We asked the crowd to assess the correctness of 100 conditional negative statements (3 annotations per statement), chosen randomly. To make it easier for annotators who are unfamiliar with RDF  triples\footnote{Especially because of the triple-pattern condition.}, we manually converted them into natural language statements, for example ``\textit{Bing Crosby did not play any keyboard instruments}''. Results show that 57\% were correct, 23\% incorrect, and 20\% were uncertain. The standard deviation of this task is 0.24 and the percentage of queries with full agreement is 18\%.

\noindent
\textbf{Usefulness.} For every entity, we showed 3 annotators 2 sets of top-3 negative statements: a grounded and universally negative statements set and a conditional negative statement set, and asked them to choose the one with more interesting information. Results are shown in Table~\ref{tab:conditionaluse}. The conditional statements were chosen 45 percentage points more than the grounded and universally negative statements. The standard deviation of this task is 0.22 and the percentage of queries with full agreement is 21\%. The significant out-performance of the conditional class over the other two classes is that it encapsulates them. Without losing the information from the original result set, lifting summarizes negations in meaningful manner, at the same time, allowing more diverse statements to be displayed in a top-k set. An example is shown in Table~\ref{tab:leolifting}, with entity $e=$\textit{Leonardo Dicaprio}, and its top-3 results. Even though he is one of the most accomplished actors in the world, unlike many of his peers, he never attempted directing any kind of creative work (films, plays, television shows, etc..).

\begin{table}
  \caption{Usefulness of conditional negative statements.}
  \centering
  \label{tab:conditionaluse}
  \begin{tabular}{l|l}
    \toprule
        \multicolumn{1}{c}{\textbf{Preferred}} & 
        \multicolumn{1}{c}{\textbf{(\%)}}\\
            \midrule
    \multicolumn{1}{l}{Conditional negative statements} & \multicolumn{1}{c}{\textbf{70}}\\
    \multicolumn{1}{l}{Grounded and universally negative statements} & \multicolumn{1}{c}{25}\\
    \multicolumn{1}{l}{Either or neither} & \multicolumn{1}{c}{5}\\
    \bottomrule
  \end{tabular}
\end{table}

\begin{table*}
  \caption{Top-3 negative statements about \textit{Leonardo Dicaprio}, before and after lifting.}
  \label{tab:leolifting}
  \centering
  \scalebox{0.9}{
  \begin{tabular}{l|l}
    \toprule
    \multicolumn{1}{c}{\textbf{Negative statements}} & \multicolumn{1}{c}{\textbf{Conditional negative statements}}\\
    \midrule
$\neg$(occupation; film director) & $\neg\exists o$ (occupation; $o$) ($o$; subclass of; director)\\
$\neg$(occupation; theater director) & $\neg \exists x$(spouse; x) \\
$\neg$(occupation; television director) & $\neg \exists x$(child; x) \\
    \bottomrule
  \end{tabular}}
  \end{table*}


%% file: extrinsic.tex
\section{Extrinsic Evaluation}
\label{sec:extrinsicevaluation}
We highlight the relevance of negative statements for:
\begin{itemize}
    \item Entity summarization on Wikidata.
    \item Decision support with hotel data from Booking.com.
    \item Question answering on various structured search engines.
\end{itemize}

\subsection{Entity Summarization}
In this experiment we analyze whether mixed positive-negative statement set can compete with standard positive-only statement sets in the task of entity summarization. In particular, we want to show that the addition of negative statements will \textit{increase the descriptive power} of structured summaries.

We collected 100 Wikidata entities from 3 diverse types: 40 people, 30 organizations (including publishers, financial institutions, academic institutions, cultural centers, businesses, and more), and 30 literary works (including creative work like poems, songs, novels, religious texts, theses, book reviews, and more). On top of the negative statements that we infered, we collected relevant positive statements about those entities.\footnote{We defined a number of common/useful properties to each of  type, e.g., for people, ``position held''is a relevant property for positive statements.} We then computed for each entity \textit{e} a sample of 10 positive-only statements, and a mixed set of 7 positive and 3 \textit{correct}\footnote{We manually checked the correctness of these negative statements.} negative statements, produced by the peer-based method. We relied on peering using Wikipedia embeddings~\cite{wikipedia2vec}. Annotators were then asked to decide which set contains more new or unexpected information about \textit{e}. More particularly, for every entity, we asked workers to assess the sets (flipping the position of our set to avoid biases), leading to a total number of 100 tasks for 100 entities. We collected 3 opinions per task. Overall results show that mixed sets with negative information were preferred for 72\% of the entities, sets with positive-only statements were preferred for 17\% of the entities, and the option ``both or neither'' was chosen for 11\% of the entities. Table~\ref{tab:posneg} shows results per each considered type.  The standard deviation is 0.24, and the percentage of queries with full agreement is 22\%.
Table~\ref{tab:won} shows three diverse examples. The first one is \textit{Daily Mirror}. One particular noteworthy negative statement in this case is that the newspaper is not owned by the ``\textit{News U.K.}'' publisher which owns a number of other \textit{British} newspapers like \textit{The Times, The Sunday Times, and The Sun}. The second entity is \textit{Peter the Great} who died in \textit{Saint Petersburg} and not \textit{Moscow}, and who did not receive the \textit{Order of St Alexander Nevsky} which was first established by his wife, a few months after his death. And the third entity is \textit{Twist and Shout}. Although it is a known song by \textit{The Beatles}, they were \textit{not} its composers, writers, nor original performers.

\begin{table*}
  \caption{Results for the entities \textit{Daily Mirror}, \textit{Peter the Great}, and \textit{Twist and Shout}.}
  \label{tab:won}
    \center
    \begin{centering}
  \scalebox{0.9}{
  \begin{tabular}{c|c}
    \toprule
      \multicolumn{2}{c}{\textbf{Daily Mirror}}\\
          \midrule
    \multicolumn{1}{c}{\textbf{Pos-only}} & \multicolumn{1}{c}{\textbf{Pos-and-neg}}\\
    \midrule 
(owned by; Reach plc) & \bf{\textit{$\neg$(newspaper format; broadsheet)}}\\
(newspaper format; tabloid) & (newspaper format; tabloid)\\
(country; United Kingdom) & \bf{\textit{$\neg$(country; U.S.)}}\\
(language of work or name; English) & (language of work or name; English)\\
(instance of; newspaper) & \bf{\textit{$\neg$(owned by; News U.K.)}}\\
... & ...\\
\midrule
      \multicolumn{2}{c}{\textbf{Peter the Great}}\\
          \midrule
    \multicolumn{1}{c}{\textbf{Pos-only}} & \multicolumn{1}{c}{\textbf{Pos-and-neg}}\\
    \midrule 
(military rank; general officer) & (military rank; general officer)\\
(owner of; Kadriorg Palace) & (owner of; Kadriorg Palace)\\
(award; Order of the Elephant) & \bf{\textit{$\neg$(place of death; Moscow)}}\\
(award; Order of St. Andrew) & (award; Order of St. Andrew)\\
(father; Alexis of Russia) & \bf{\textit{$\neg$(award; Knight of the Order of St. Alexander Nevsky)}}\\
... & ...\\
\midrule
      \multicolumn{2}{c}{\textbf{Twist And Shout}}\\
          \midrule
    \multicolumn{1}{c}{\textbf{Pos-only}} & \multicolumn{1}{c}{\textbf{Pos-and-neg}}\\
    \midrule 
(composer; Phil Medley)& \bf{\textit{$\neg$(composer; Paul McCartney)}}\\
(performer; The Beatles) & (performer; The Beatles)\\
(producer; George Martin) & \bf{\textit{$\neg$(composer; John Lennon)}}\\
(instance of; musical composition) & (instance of; musical composition)\\
(lyrics by; Phil Medley) & \bf{\textit{$\neg$(lyrics by; Paul McCartney)}}\\
... & ...\\
    \bottomrule
  \end{tabular}
  }
  \end{centering}
  \end{table*}
  
\begin{table*}
  \caption{Positive-only vs.\ positive and negative statements.}
  \label{tab:posneg}
  \center
  \begin{centering}
  \scalebox{0.99}{
  \begin{tabular}{l|l|l|l}
    \toprule
        \multicolumn{1}{c}{\textbf{Preferred Choice}} & 
        \multicolumn{1}{c}{\bf{Person} \textbf{(\%)}} & \multicolumn{1}{c}{\bf{Organization} \textbf{(\%)}} & \multicolumn{1}{c}{\bf{Literary work} \textbf{(\%)}}\\
            \midrule
    \multicolumn{1}{l}{Pos-and-neg} & \multicolumn{1}{c}{\textbf{71}} &\multicolumn{1}{c}{\textbf{77}}&\multicolumn{1}{c}{\textbf{66}}\\
    \multicolumn{1}{l}{Pos-only} & \multicolumn{1}{c}{22} &  \multicolumn{1}{c}{10} & \multicolumn{1}{c}{17} \\
    \multicolumn{1}{l}{Both or neither} & \multicolumn{1}{c}{7} & \multicolumn{1}{c}{13} & \multicolumn{1}{c}{17}\\
    \bottomrule
  \end{tabular}
  }
  \end{centering}
\end{table*}

In this experiment, we showed that adding negative statements to a set of positive statements increases its quality, and for that, we chose a split of 7 positive and 3 negative statements for top-10 results. One may wonder whether that is actually the best proportion. This motivates another analysis, \textit{finding out the portion of negative statements to be added to a positive top-k set of statements that maximizes the relevance gain} (i.e., nDCG). We used the annotators' assessment of relevancy of individual positive and negative statements. 
 We then compiled them as sets of top-k results with different k values and different portions of negative statements. The decision of adding a certain negative statement should respect the constraint of not decreasing the relevance gain (i.e., nDCG) of the currently chosen top-k results. We calculated the ideal ratio of positive to negative statements for k results. The ideal portion of negative statements within top-k statements about entity \textit{e} was obtained for k=3, 5, 10, and 20. For a set of top-3 or top-5 statements, 1 negative statement is ideal, for 10 statements, 2 are ideal, and for 20, 5 are ideal.

\subsection{Decision Support}
Negative statements are highly important also in specific domains. In online shopping, characteristics not possessed by a product, such as the \textit{IPhone 7} not having a headphone jack, are a frequent topic highly relevant for decision making. The same applies to the hospitality domain: the absence of features such as free WiFi or gym rooms are important criteria for hotel bookers, although portals like Booking.com currently only show (sometimes overwhelming) positive feature sets.

To illustrate this, based on a comparison of 1.8k hotels in India, as per their listing on Booking.com, using the peer-based method, we inferred useful negative features. For peering, we considered all other hotels in India, and for ranking, we computed peer frequencies (\textit{PEER}). We then used crowdsourcing over the results of 100 hotels. We asked annotators to check two sets of features about a given hotel, one set containing 5 random 
positive-only features, and one set containing a mix of 3 positive and 2 negative features. Their task was to choose which set of features will help them more in deciding whether to stay in this hotel or not. They can choose one of the sets, or both. For every hotel, we request 3 annotators.

Table~\ref{tab:hotelsnumber} shows that sets with negative features were chosen 16 percentage points more than the positive-only sets. The standard deviation of this task is 0.22 and the percentage of queries with full agreement is 28\%. Table~\ref{tab:hotels} shows three hotels with useful negative features. Although the \textit{Hotel Asia The Dawn} lists 64 positive features, negative information such as that it does not offer air conditioning and free Wifi may give important clues for decision making.
\begin{table}
  \caption{Usefulness of hotel features.}
  \centering
  \label{tab:hotelsnumber}
  \begin{tabular}{l|l}
    \toprule
        \multicolumn{1}{c}{\textbf{Preferred Choice}} & 
        \multicolumn{1}{c}{\textbf{(\%)}}\\
            \midrule
    \multicolumn{1}{l}{Pos-and-neg} & \multicolumn{1}{c}{\textbf{54}}\\
    \multicolumn{1}{l}{Pos-only} & \multicolumn{1}{c}{38}\\
    \multicolumn{1}{l}{Either or neither} & \multicolumn{1}{c}{8}\\
    \bottomrule
  \end{tabular}
\end{table}

\begin{table*}
\caption{Negative statements for hotels in India.}
\centering
\scalebox{0.9}{\begin{tabular}{lcl}
\toprule
\bf{Hotel} & \bf{Number of positive features} & \bf{Top-3 negative features} \\ \midrule
The Sultan Resort & 106 & $\neg$ Parking; $\neg$ Fan; $\neg$ Newspapers\\
Vista Rooms at Mount Road & 28 & $\neg$ Room service; $\neg$ Food \& Drink; $\neg$ 24-hour front desk\\
Hotel Asia The Dawn & 64 & $\neg$ Air conditioning; $\neg$ Free Wifi; $\neg$ Free private parking\\
\bottomrule
\end{tabular}}
\label{tab:hotels}
\end{table*}

\begin{figure*}
 \caption{Extrinsic use-case: decision support on hotel data.}
 \centering
\includegraphics[width=0.85\textwidth]{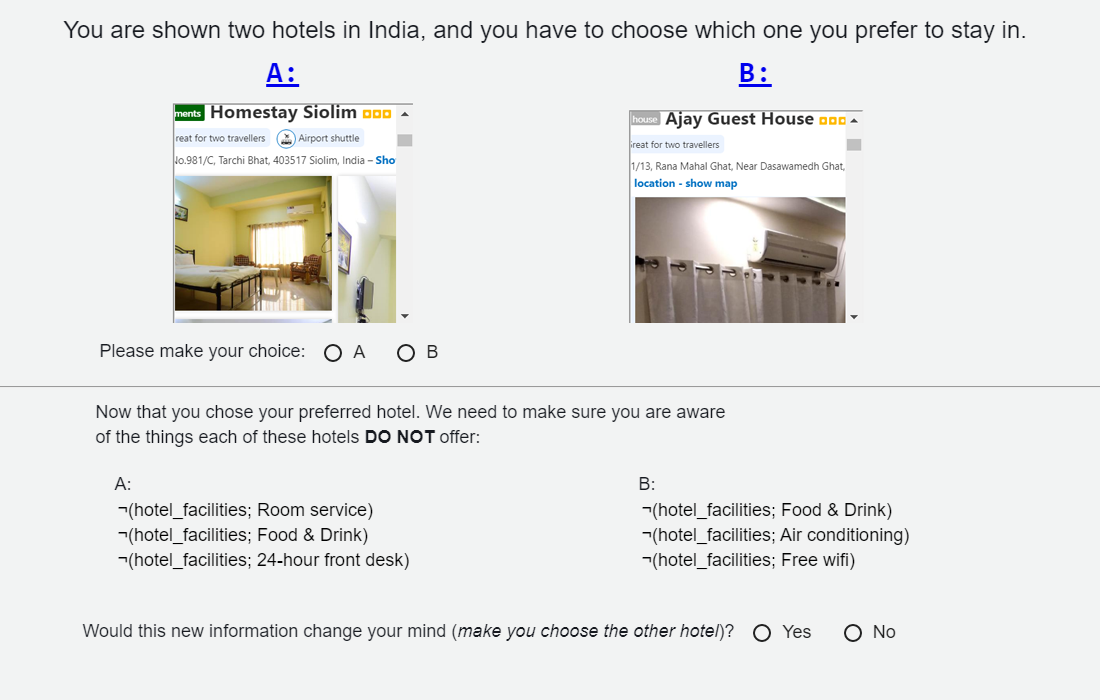}
\label{fig:hotel}
\end{figure*}

Moreover, we collected 20 pairs of hotels from the same dataset, and showed every pair's Booking.com pages to 3 annotators. We asked them to choose the better hotel for them. Then we showed them negative features about the pair, and asked them whether this new information would change their mind on their initial decision. A screenshot of the task is shown in Figure~\ref{fig:hotel}. 42\% changed their pick after negative features were revealed. The standard deviation on this task is 0.15. The full agreement of the 3 annotators on \textit{changing the hotel after negative features were revealed} is 35\%. The full agreement of annotators \textit{choosing the same hotel at the end of the task} is 30\%. The latter agreement measure disregard whether they have changed their decision or stayed with their initial choice.

\subsection{Question Answering}
In this experiment, we compared the results to negative questions over a diverse set of sources. We manually compiled 20 questions that involve negation, such as \emph{``Actors without Oscars''}\footnote{Sample textual queries: ``actors with no Oscars'', ``actors with no spouses'', ``film actors who are not film directors'', ``football players with no Ballon d'Or'', ``politicians who are not lawyers''.}. We compared them over four highly diverse sources: Google Web Search (increasingly returning structured answers from the Google knowledge graph~\cite{GKG}), WDAqua~\cite{Diefenbach2017} (an academic state-of-the-art KBQA system), the Wikidata SPARQL endpoint~\footnote{\url{https://query.wikidata.org/}} (direct access to structured data), and our peer-based method. For Google Web Search and WD\-Aqua, we submitted the queries in their textual form, and considered answers from Google if they come as structured knowledge panels. For Wikidata and peer-based inference, we transform the queries into SPARQL queries\footnote{sample SPARQL queries: \url{https://w.wiki/A6r}, \url{https://w.wiki/9yk}, \url{https://w.wiki/9yn}, \url{https://w.wiki/9yp}, \url{https://w.wiki/9yq}}, which we either fully executed over the Wikidata endpoint, or executed the positive part over the Wikidata endpoint, while evaluating the negative part over a dataset produced by our peer-based inference method. Note that all queries were safe, since they were designed to always asks for a class of entities (e.g., entities of occupation actor) that do not satisfy a certain property (e.g., having won the Oscar), which was captured via SPARQL MINUS with a shared variable. For each method, we then self-evaluated the number of results, the correctness and relevance of the (top-5) results. All methods were able to return highly correct statements, yet Google Web Search and WDAqua return no results for 18 and 16 of the queries, respectively.

We continued the assessment over a sample of 5 queries. Wikidata SPARQL returned by far the highest number of results, 250k on average, yet did not perform ranking, thus returned results that are hardly relevant (e.g., a local Latvian actor to the Oscar question). The peer-based inference outperforms it by far in terms of relevance (72\% vs.\ 44\% for Wikidata SPARQL). We point out that although Wikidata SPARQL results appear highly correct, this has no formal foundation, due to the absence of a stance of OWA KBs towards negative knowledge. For example, most actors or people did \textit{not} win Oscars, which makes 99.99\% of the entities returned by Wikidata's SPARQL query correct, even under the OWA.


%% file: resources.tex
\section{Resources}
\label{sec:datasets}
\noindent
\textbf{Negative Statement Datasets for Wikidata.\ }
\label{sec:dataset}
We publish the first 
datasets that contain dedicated \emph{useful} negative statements about entities in Wikidata: (i) Peer-based and order-oriented inference data: 14m negative statements about popular 600k entities from various types, (ii) release the mturk-annotated on the correctness of 1k negative statements of Section~\ref{subsec:similarityexp}, and (iii) 40k ordered set of peers introduced in Section~\ref{sub:temporalexperiments}.

\noindent
\textbf{Open-source Code.\ } We publish our code for peer-based inference, so others can execute it on their own datasets \footnote{\url{https://github.com/HibaArnaout/usefulnegations}}.

\noindent
\textbf{Demo.\ }
\label{sec:demo}
A web-based platform, Wikinegata~\cite{ArnaoutRWP21,arnaout2021negative} for browsing useful negations about Wikidata entities, is available at: \url{https://d5demos.mpi-inf.mpg.de/negation/}.

\begin{figure*}
 \caption{Interface for Wikinegata - useful negative statements about \textit{Leonardo DiCaprio}.}
 \centering
\includegraphics[width=0.9\textwidth]{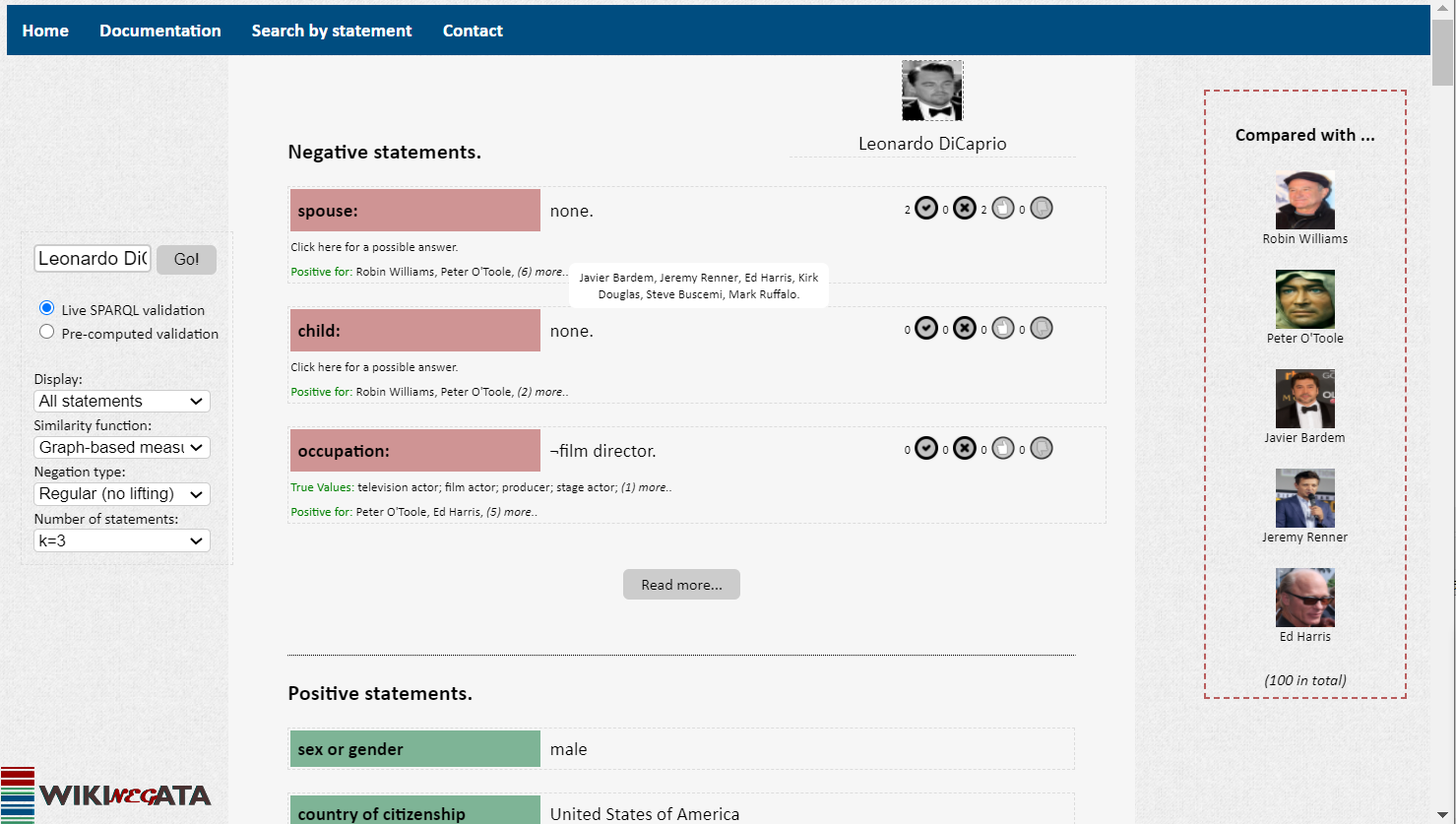}
\label{fig:demo}
\end{figure*}

\noindent
A screenshot is shown in Figure~\ref{fig:demo}. 

All experimental material related to this paper can be found on a dedicated webpage\footnote{\url{https://www.mpi-inf.mpg.de/departments/databases-and-information-systems/research/knowledge-base-recall/interesting-negations-in-kbs}}.

%% file: discussion.tex
\section{Discussion}
\label{sec:discussion}
\subsection{Quality Considerations}

\noindent
\textbf{The CWA on the Semantic Web.\ }
Negation has traditionally been avoided on the Semantic Web, as it challenges the vision that anyone can state anything, without risking logical conflicts. In the present work, we showed that enriching KBs with useful negative statements is beneficial in use cases such as entity summarization and consumer decision making. In order to compile a set of likely correct negative statements about an entity, we assumed the CWA in parts of the KBs, namely within peer groups, and in the case of 
grounded
%
negative statements, with the additional requirement that there is at least one other positive statement for the same entity-property pair. Although this approach outperforms other techniques, like embedding-based KB completion, inferences may still be incorrect. While correctness can be tuned to some extend by sacrificing recall (e.g., requiring very high thresholds on \textit{PEER} and \textit{PIVO}), errors are still possible. It is therefore advised to show 
candidate statements from automatic inference to KB curators
for final assessment~\cite{RECOIN}.

\noindent
\textbf{Real-world Changes and KB Maintenance.\ } 
Due to real-world changes or new information added to the KB, some of the negative statements already inferred might become incorrect. For instance, \textit{DiCaprio} has won his first \textit{Oscar} in 2016. After the year 2016, the negative statement \term{$\neg$(DiCaprio; award; Oscar)} is no longer correct. Negative statements should therefore be timestamped, and ideally, additions of positive statements should automatically trigger updates of validity end-point timestamps.

\noindent
\textbf{Class Hierarchies.\ }
Some incorrect negations can be detected by help of subsumption checks (rdfs:subClassOf). For example, the presented method might incorrectly infer the statement \term{$\neg$(Douglas Adams; occupation; author)}, which contradicts the two positive assertions that \textit{Douglas Adams is a writer}, and \textit{writer} is a subclass of \textit{author}. One could detect such contradictions by use of a generic ontology reasoner like Protégé, or implement custom checks. For our specific use case of negative inference at scale, we found that checks focused on one or two hops in the class hierarchy capture a significant proportion of these errors. For KBs at the scale of Wikidata, one could precompute prominent subsumptions, and build these checks into the methodology (e.g., triggering a check for presence of ``occupation-writer'' whenever ``occupation-not author'' is inferred).

Subsumption similarly also affects properties: the relation \textit{CEO} (between a company and a person) is a subproperty of \textit{employee}, and as such, subject-object-pairs present for the former should not appear as negations for the latter.



\noindent
\textbf{Modelling and Constraint Enforcement.\ } 
Some inferred negations are incorrect due to modelling issues, resulting in inconsistencies. An example is \textit{Dijkstra} and the negative statement that his field of work is \textit{not} \textit{Computer Science}, and \textit{not} \textit{Information Technology}, while he has the positive value \textit{Informatics}, which is arguably near-synonymous, yet in the Wikidata taxonomy, the two represent independent concepts, two hops apart. Some other incorrect negative statements could be due to a lack of constraints. For instance, for most businesses, the \textit{headquarters location} property is completed using \textit{cities}, but for \textit{Siemens} in Wikidata, the \textit{building} is added instead (\textit{Palais Ludwig Ferdinand}), making our inferred statement \term{$\neg$(Siemens; headquarters location; Munich)} incorrect. Although Wikidata encourages editors to use cities for the  \textit{headquarters location} property and advise them to use another property for specific buildings, it has not been automatically enforced yet.\footnote{\url{https://www.wikidata.org/wiki/Property:P159}}

\subsection{Discovering Relevant Lifting Aspects}
For inferring conditional negative statements, the lifting aspects we used in this paper have been manually defined (see Table~\ref{tab:aspects}). For instance, if the grounded negative statements to be lifted describe educational institutions, then the aspects that make sense are the location of the institution (\textit{U.S., Germany, Japan, etc.}) and its type (\textit{public, private, research, etc.}). This 
does not scale well when the KB contains thousands of properties with thousands of possible aspects. Automatically discovering these aspects would improve the quality of conditional negative statements. A good start is the work in~\cite{oren2006extending}. 
An aspect is described as an important characteristic of an entity. For example, for a book, the number of pages is not an important aspect, but genre is.
This work introduces aspect ranking metrics such as object cardinality: a \textit{good} predicate (e.g., genre) has a finite list of values to choose from (e.g., comedy, thriller, romance). Unsuitable predicates using this metric would be the predicate \textit{number of pages} or \textit{publication date}. In addition, the AMIE system~\cite{AMIE,AMIEP} mines rules on millions of triples, and is specifically tailored to support open-world KBs. It can discover, for example, that \textit{musicians that are influenced by each other often play the same instrument.} The \textit{instrument} can be directly used as an aspect for lifting grounded negative statements (with predicate \textit{influenced by}) about a \textit{musician-entity}. In particular, \textit{musician x} (a pianist), is not influenced by anyone who plays the guitar, or more surprisingly not influenced by anyone who plays the piano (if that is the case). We consider this to be a promising research direction. It is worth exploring and improving the ideas in Section~\ref{sec:restricted} further.

\subsection{Entity Prominence and Class Specificity}

\noindent
\textbf{Negations in the Long Tail.\ }
Our method builds on the assumption that peer entities are available, 
for which we have sufficient data.
%
For long-tail entities, both assumptions may be challenged. For entities with extremely little positive information (e.g., \url{https://www.wikidata.org/wiki/Q97355589}, for which only first name, last name, and gender are known), it is not possible to identify relevant peers, and hence, our method is not applicable. Low amounts of positive information on peers, in contrast, can be better compensated.
Since our method is mainly concerned with finding the most interesting candidates for negation, absolute frequencies are not important, as long as it is possible to find a reasonable difference in frequencies among peers (i.e., not every positive statement appearing only once). If there is interest to put emphasis on specific facts, one could also 
adjust
the ranking algorithms, e.g., giving ``citizenship'' negations a 
boost in the ranking.

\begin{table*}
  \caption{Negations across classes of Wikidata entities.}
  \centering
  \label{tab:differenttypes}
    \scalebox{0.8}{\begin{tabular}{l|l|l|l}
    \toprule
        \multicolumn{1}{c}{\textbf{Class}} & 
        \multicolumn{1}{c}{\textbf{Number of entities}} &
        \multicolumn{1}{c}{\textbf{3 most frequent negated properties}} &  \multicolumn{1}{c}{\textbf{Sample entities}}\\
            \midrule
    \multicolumn{1}{l}{Book} & \multicolumn{1}{c}{8k} & \multicolumn{1}{l}{author, genre, publisher} & Fahrenheit 451, Little Birds\\
    \multicolumn{1}{l}{Person} & \multicolumn{1}{c}{500k} & \multicolumn{1}{l}{spouse, child, occupation} & Elon Musk, Oprah Winfrey\\
    \multicolumn{1}{l}{Country} & \multicolumn{1}{c}{199} & \multicolumn{1}{l}{diplomatic relation, member of, language used} & Germany, China\\
     \multicolumn{1}{l}{Primary school} & \multicolumn{1}{c}{14k} & \multicolumn{1}{l}{instance of, heritage designation, country} & Deutsche Schule Helsinki, Saint Joseph school\\
     \multicolumn{1}{l}{Film} & \multicolumn{1}{c}{26k} & \multicolumn{1}{l}{cast member, genre, screenwriter} & Taxi Driver, Inception\\    
     \multicolumn{1}{l}{Building} & \multicolumn{1}{c}{28k} & \multicolumn{1}{l}{architect, instance of, heritage designation} & NY Times Building, White House\\ 
     \multicolumn{1}{l}{Organizations} & \multicolumn{1}{c}{22k} & \multicolumn{1}{l}{headquarters location, instance of, country} & World Trade Organization, BBC\\
     \multicolumn{1}{l}{Musical group} & \multicolumn{1}{c}{8k} & \multicolumn{1}{l}{instance of, record label, genre} & Coldplay, Jonas Brothers\\ 
     \multicolumn{1}{l}{Business} & \multicolumn{1}{c}{20k} & \multicolumn{1}{l}{parent organization, headquarters location, industry} & Nokia, Facebook\\ 
     \multicolumn{1}{l}{Scientific journal} & \multicolumn{1}{c}{5k} & \multicolumn{1}{l}{main subject, editor, publisher} & Journal of Web Semantics, Nature\\
     \multicolumn{1}{l}{Literary work} & \multicolumn{1}{c}{24k} & \multicolumn{1}{l}{author, composer, lyrics by} & Diary of Anne Frank, Don Quixote\\
    \bottomrule
  \end{tabular}}
\end{table*}

\noindent
\textbf{Negations for Different Classes.\ }
In practice, we have applied our method on 11 diverse classes of entities: people, literary works, organizations, businesses, scientific journals, countries, buildings, musical groups, primary schools, books, and films. We have observed that within each class, interesting negations often cover the same properties. For instance, for people, interesting negative knowledge is mostly about awards, occupations, education, and 
family.
We show 
statistics on frequent properties for every class of entities in Table~\ref{tab:differenttypes}. We do \textit{not} filter nor assign weights for certain properties per class. The relative frequency metric takes care of prioritizing which property's negation \textit{makes sense} in every class. For \textit{people}, the reported properties are fairly {general} and not 
tied to
specific subsets of this very large class. For instance, for sports figures, \textit{member of sports team} is the most frequent property, and for politicians, \textit{position held} is the dominating property.

We notice that negations for small classes, such as buildings and literary works,
have a higher correctness ratio
than larger classes, such as people. Entities of type \textit{person} have 3 times more possible properties to fill than entities of type \textit{book}. Given a book (e.g., \textit{Orientalism}), a handful of properties and property-object pairs could be added and the information about the entity is considered near-complete (e.g., \textit{main subject}, \textit{author}, \textit{genres}, \textit{publisher}, and \textit{language}). 
In contrast,
for a person (e.g., \textit{Joe Rogan}), the entity requires a greater effort and/or larger information sources to be considered complete (e.g.,  \textit{occupation}, \textit{education}, \textit{residence}, \textit{birth place}, \textit{citizenship}, \textit{sport}, \textit{religion}, and many more). On the other hand, larger classes offer richer and more diverse possibilities for \textit{interesting} negations. A result set for a person often covers a wider range of topics, such as personal information, professional achievements, relations with other people. A result set for a book is less diverse, often negating the same property repeatedly with different objects.\\

%% file: conclusion.tex
\section{Conclusion \& Future Directions}

This article has made the first comprehensive case for explicitly materializing useful negative statements in KBs. We have introduced a statistical inference approach on retrieving and ranking candidate negative statements, based on expectations set by highly related peers. We have also released several resources to encourage further research.

In future work we would like to explore a number of research directions:
\begin{enumerate}
    \item Missing vs negative statements: How to maximize trade-ability between fewer highly correct statements, and larger sets of interesting negation candidates.
    \item Mining complex negations: Our focus was on simple - grounded and universal - negation, with a hint at more complex conditional statements. It is open to extend that to (i) automatically finding aspects, (ii) further joins \textit{``did not study at a university which was graduating any Nobel prize winner''}, (iii) negation on sets of entities instead of entity-centric \textit{``no African country has hosted any Olympic games''}, etc.
    \item Exploring how textual information extraction of implicit negations can boost negation coverage, e.g., statements like \textit{``Theresa May is an only child.''} (corresponding to \term{$\neg \exists x$(sibling; x)}), or \textit{``George Washington had no formal education.''} (corresponding to \term{$\neg \exists x$(educated at; x)}).
    \item Exploiting the ontology that comes with the KB to improve the correctness of inferred negations by making use of constraints like class and property subsumption.
\end{enumerate}

\section*{Acknowledgement}
\noindent This work is supported by the German Science Foundation (DFG: Deutsche Forschungsgemeinschaft) by grant 4530095897: ``Negative Knowledge at Web Scale''.

%% file: main.bbl
\begin{thebibliography}{10}
\providecommand{\url}[1]{\texttt{#1}}
\providecommand{\urlprefix}{URL }
\providecommand{\doi}[1]{https://doi.org/#1}

\bibitem{DBLP:books/aw/AbiteboulHV95}
Abiteboul, S., Hull, R., Vianu, V.: Foundations of Databases. Addison-Wesley
  (1995)

\bibitem{DBLP:journals/amai/AnalytiADP13}
Analyti, A., Antoniou, G., Viegas~Dam{\'{a}}sio, C., Pachoulakis, I.: A
  framework for modular {ERDF} ontologies. AMAI  (2013)

\bibitem{Analyti04negationand}
Analyti, A., Antoniou, G., Viegas~Dam{\'{a}}sio, C., Wagner, G.: Negation and
  negative information in the {W3C} {Resource} {Description} {Framework}. AMCT
  (2004)

\bibitem{arnaoutjws2018}
Arnaout, H., Elbassuoni, S.: Effective searching of {RDF} knowledge graphs. JWS
   (2018)

\bibitem{negationakbc}
Arnaout, H., Razniewski, S., Weikum, G.: Enriching knowledge bases with
  interesting negative statements. In: AKBC (2020)

\bibitem{ArnaoutRWP21}
Arnaout, H., Razniewski, S., Weikum, G., Pan, J.Z.: Wikinegata: a knowledge
  base with interesting negative statements. PVLDB  (2021)

\bibitem{arnaout2021negative}
Arnaout, H., Razniewski, S., Weikum, G., Pan, J.Z.: Negative knowledge for
  open-world wikidata. In: The Web Conference (2021)

\bibitem{DBPEDIA}
Auer, S., Bizer, C., Kobilarov, G., Lehmann, J., Cyganiak, R., Ives, Z.:
  {DBpedia}: A nucleus for a web of open data. In: ISWC (2007)

\bibitem{logichandbook}
Baader, F., Calvanese, D., Mcguinness, D., Nardi, D., F.~Patel-Schneider, P.:
  The Description Logic handbook. Cambridge University Press (2007)

\bibitem{RECOIN}
Balaraman, V., Razniewski, S., Nutt, W.: Recoin: relative completeness in
  {Wikidata}. In: Wiki Workshop at WWW (2018)

\bibitem{Bast}
Bast, H., Buchhold, B., Haussmann, E.: Relevance scores for triples from
  type-like relations. In: SIGIR (2015)

\bibitem{transE}
Bordes, A., Usunier, N., Garcia-Duran, A., Weston, J., Yakhnenko, O.:
  Translating embeddings for modeling multi-relational data. In: NIPS (2013)

\bibitem{PRENEX}
Bärbäntan, I., Potolea, R.: Exploiting word meaning for negation
  identification in electronic health records. In: AQTR (2014)

\bibitem{KEhealth}
Bärbäntan, I., Potolea, R.: Towards knowledge extraction from electronic
  health records - automatic negation identification. In: IFMBE (2014)

\bibitem{Calvanese2007}
Calvanese, D., De~Giacomo, G., Lembo, D., Lenzerini, M., Rosati, R.: Tractable
  reasoning and efficient query answering in {Description Logics}: The
  {DL-Lite} family. Journal of Automated Reasoning  (2007)

\bibitem{CHAPMAN}
Chapman, W., Bridewell, W., Hanbury, P., Cooper, G., Buchanan, B.: A simple
  algorithm for identifying negated findings and diseases in discharge
  summaries. Journal of Biomedical Informatics  (2001)

\bibitem{extendingnegex}
Chapman, W., Hillert, D., Velupillai, S., Kvist, M., Skeppstedt, M., Chapman,
  B., Conway, M., Tharp, M., Mowery, D., Deleger, L.: Extending the {NegEx}
  lexicon for multiple languages. Studies in health technology and informatics
  (2013)

\bibitem{cruzdiaz}
Cruz~D{\'\i}az, N.P.: Detecting negated and uncertain information in biomedical
  and review texts. In: RANLP (2013)

\bibitem{DBLP:conf/semweb/DarariPN15}
Darari, F., Prasojo, R.E., Nutt, W.: Expressing no-value information in {RDF}.
  In: ISWC (2015)

\bibitem{Diefenbach2017}
Diefenbach, D., Singh, K., Maret, P.: {WDAqua-core0}: A question answering
  component for the research community. In: ESWC (2017)

\bibitem{DuPa2015}
Du, J., Pan, J.Z.: {Rewriting-based instance retrieval for negated concepts in
  Description Logic ontologies}. In: ISWC (2015)

\bibitem{ernst2015knowlife}
Ernst, P., Siu, A., Weikum, G.: {Knowlife}: a versatile approach for
  constructing a large knowledge graph for biomedical sciences. BMC
  bioinformatics  (2015)

\bibitem{erxleben2014introducing}
Erxleben, F., G{\"u}nther, M., Kr{\"o}tzsch, M., Mendez, J., ,
  Vrande{\v{c}}i{\'c}, D.: Introducing {Wikidata} to the linked data web. In:
  ISWC (2014)

\bibitem{Fader2014}
Fader, A., L., Z., O., E.: Open question answering over curated and extracted
  knowledge bases. In: KDD (2014)

\bibitem{FHPPW2006}
Flouris, G., Huang, Z., Pan, J.Z., Plexousakis, D., Wache, H.: Inconsistencies,
  negations and changes in ontologies. In: AAAI (2006)

\bibitem{galarraga2017predicting}
Gal{\'a}rraga, L., Razniewski, S., Amarilli, A., Suchanek, F.M.: Predicting
  completeness in knowledge bases. In: WSDM (2017)

\bibitem{AMIEP}
Gal\'{a}rraga, L., Teflioudi, C., Hose, K., Suchanek, F.M.: Fast rule mining in
  ontological knowledge bases with {AMIE}+. VLDB  (2015)

\bibitem{AMIE}
Galárraga, L.A., Teflioudi, C., Hose, K., Suchanek, F.M.: {AMIE}: association
  rule mining under incomplete evidence in ontological knowledge bases. In: WWW
  (2013)

\bibitem{ghoshSWJ}
Ghosh, S., Razniewski, S., Weikum, G.: Uncovering hidden semantics of set
  information in knowledge bases. JWS  (2020)

\bibitem{Goldin03learningto}
Goldin, I., Chapman, W.: Learning to detect negation with {``Not''} in medical
  texts. In: SIGIR (2003)

\bibitem{ho2018rule}
Ho, V., Stepanova, D., Gad-Elrab, M., Kharlamov, E., Weikum, G.: Rule learning
  from knowledge graphs guided by embedding models. In: ISWC (2018)

\bibitem{huang2019contextual}
Huang, S., Liu, J., Korn, F., Wang, X., Wu, Y., Markowitz, D., Yu, C.:
  Contextual fact ranking and its applications in table synthesis and
  compression. In: KDD (2019)

\bibitem{NDCG}
J\"{a}rvelin, K., Kek\"{a}l\"{a}inen, J.: Cumulated gain-based evaluation of
  {IR} techniques. Trans. Inf. Syst.  (2002)

\bibitem{ANION}
Jiang, L., Bosselut, A., Bhagavatula, C., Choi, Y.: ``{I'm} not mad'':
  commonsense implications of negation and contradiction. In: NAACL-HLT (2021)

\bibitem{AKB}
Karagiannis, G., Trummer, I., Jo, S., Khandelwal, S., Wang, X., Yu, C.: Mining
  an "anti-knowledge base" from {Wikipedia} updates with applications to fact
  checking and beyond. PVLDB  (2019)

\bibitem{NAGA}
Kasneci, G., Suchanek, F.M., Ifrim, G., Ramanath, M., Weikum, G.: {NAGA}:
  Searching and ranking knowledge. In: ICDE (2008)

\bibitem{AMIE3}
Lajus, J., Gal{\'a}rraga, L., Suchanek, F.: Fast and exact rule mining with
  {AMIE} 3. In: ESWC (2020)

\bibitem{MKGGB2018}
Malyshev, S., Kr{\"{o}}tzsch, M., Gonz{\'{a}}lez, L., Gonsior, L., Bielefeldt,
  A.: Getting the most out of {Wikidata}: semantic technology usage in
  {Wikipedia’s} knowledge graph. In: ISWC (2018)

\bibitem{mcguinness2004owl}
McGuinness, D., Van~Harmelen, F., et~al.: {OWL} web ontology language overview.
  W3C recommendation  (2004)

\bibitem{ICWA}
Minker, J.: On indefinite databases and the closed world assumption. In: 6th
  Conference on Automated Deduction (1982)

\bibitem{Morante2012}
Morante, R., Sporleder, C.: Modality and negation: An introduction to the
  special issue. Comput. Linguist.  (2012)

\bibitem{holE}
Nickel, M., Rosasco, L., Poggio, T.: Holographic embeddings of knowledge
  graphs. In: AAAI (2016)

\bibitem{oren2006extending}
Oren, E., Delbru, R., Decker, S.: Extending faceted navigation for {RDF} data.
  In: ISWC (2006)

\bibitem{ortona2018rudik}
Ortona, S., Meduri, V., Papotti, P.: {RuDiK}: rule discovery in knowledge
  bases. VLDB  (2018)

\bibitem{Pan2017}
Pan, J.Z., Calvanese, D., Eiter, T., Horrocks, I., Kifer, M., Lin, F., (Eds),
  Y.Z.: Reasoning web: logical foundation of knowledge graph construction and
  query answering. Springer (2017)

\bibitem{PPLL+2018}
Pan, J.Z., Pavlova, S., Li, C., Li, N., Li, Y., Liu, J.: Content based fake
  news detection using knowledge graphs. In: ISWC (2018)

\bibitem{global2014}
Pennington, J., Socher, R., Manning, C.: Glove: Global vectors for word
  representation. In: EMNLP (2014)

\bibitem{ponza}
Ponza, M., Ferragina, P., Chakrabarti, S.: A two-stage framework for computing
  entity relatedness in {Wikipedia}. In: CIKM (2017)

\bibitem{razniewski2017doctoral}
Razniewski, S., Balaraman, V., Nutt, W.: Doctoral advisor or medical condition:
  towards entity-specific rankings of knowledge base properties. In: ADMA
  (2017)

\bibitem{razniewski2011completeness}
Razniewski, S., Nutt, W.: Completeness of queries over incomplete databases.
  VLDB (2011)

\bibitem{razniewskilimits}
Razniewski, S., Arnaout, H., Ghosh, S., Suchanek, F.: On the limits of machine
  knowledge: Completeness, recall and negation in web-scale knowledge bases.
  VLDB  (2021)

\bibitem{Reiter78}
Reiter, R.: On Closed World Data Bases (1978)

\bibitem{RPZ2010c}
Ren, Y., Pan, J.Z., Zhao, Y.: Closed world reasoning for {OWL2} with {NBox}.
  JTST  (2010)

\bibitem{CSKB}
Safavi, T., Koutra, D.: Generating negative commonsense knowledge. arXiv
  (2020)

\bibitem{GKG}
Singhal, A.: Introducing the knowledge graph: things, not strings.
  \url{https://www.blog.google/products/search/introducing-knowledge-graph-things-not}
  (2012)

\bibitem{YAGO}
Suchanek, F., Kasneci, G., , Weikum, G.: {Yago}: A core of semantic knowledge.
  In: WWW (2007)

\bibitem{bioscope}
Szarvas, G., Vincze, V., Farkas, R., Csirik, J.: The {BioScope} corpus:
  Annotation for negation, uncertainty and their scope in biomedical texts.
  BioNLP (2008)

\bibitem{edithistory2019}
Tanon, T., Suchanek, F.: Querying the edit history of {Wikidata}. In: ESWC
  (2019)

\bibitem{gadwww}
Tran, T., Gad-Elrab, M., Stepanova, D., Kharlamov, E., Str\"{o}tgen, J.: Fast
  computation of explanations for inconsistency in large-scale knowledge
  graphs. In: WWW (2020)

\bibitem{WD}
Vrande{\v{c}}i{\'c}, D., Kr\"{o}tzsch, M.: {Wikidata}: a free collaborative
  knowledge base. CACM  (2014)

\bibitem{KGembsurvey}
Wang, Q., Mao, Z., Wang, B., Guo, L.: Knowledge graph embedding: a survey of
  approaches and applications. IEEE TKDE  (2017)

\bibitem{WPKD2020}
Wiharja, K., Pan, J.Z., Kollingbaum, M.J., Deng, Y.: Schema aware iterative
  knowledge graph completion. Journal of Web Semantics  (2020)

\bibitem{wu2014}
Wu, S., Miller, T., Masanz, J., Coarr, M., Halgrim, S., Carrell, D., Clark, C.:
  Negation's not solved: generalizability versus optimizability in clinical
  natural language processing. PloS one  (2014)

\bibitem{Yahya2016}
Yahya, M., Barbosa, D., Berberich, K., Wang, Q., Weikum, G.: Relationship
  queries on extended knowledge graphs. In: WSDM (2016)

\bibitem{wikipedia2vec}
Yamada, I., Asai, A., Shindo, H., Takeda, H., Takefuji, Y.: {Wikipedia2Vec}: an
  optimized tool for learning embeddings of words and entities from wikipedia.
  arXiv  (2018)

\bibitem{WIKIQA}
Yang, Y., Yih, W., Meek, C.: {W}iki{QA}: a challenge dataset for open-domain
  question answering. In: EMNLP (2015)

\end{thebibliography}
